\documentclass[11pt, reqno]{amsart}
\usepackage{graphicx,amsmath,amssymb,epsfig, float,xcolor}
\usepackage{amsthm,thmtools}
\usepackage{amsfonts, lscape}
\usepackage{color, caption}
\usepackage{bbm}
\usepackage{harvard,verbatim}
\usepackage{threeparttable}
\usepackage{url}
\usepackage{pdfsync}
\usepackage[latin1]{inputenc}
\usepackage{amssymb}
\usepackage{mathrsfs}
\usepackage{rotating}

\usepackage{hyperref}

\hypersetup{
    colorlinks=true,
    urlcolor=blue,
    linkcolor=blue,
    citecolor=blue
}

\theoremstyle{plain}

\newtheorem{algorithm}{Algorithm}
\newtheorem{thm}{Theorem}[section]
\newtheorem{coro}[thm]{Corollary}
\newtheorem{prop}[thm]{Proposition}
\newtheorem{lem}[thm]{Lemma}
\newtheorem{defi}[thm]{Definition}
\theoremstyle{definition}

\theoremstyle{remark}

\theoremstyle{definition}

\numberwithin{equation}{section}

\theoremstyle{definition}

\graphicspath{ {FigsEtc/} }

\providecommand\mathbb{\bf}

\let \eps\varepsilon
\let \phi\varphi

\newcommand\CC{{\mathcal C}}

\def\<#1,#2>{\left<#1,#2\right>}
\let\bar\overline
\usepackage{graphicx}
\graphicspath{ {img/} }
\input{tcilatex}

\begin{document}
\title[Vector quantile regression from theory to numerics]{Vector quantile
regression and optimal transport, from theory to numerics}
\author[G. Carlier]{Guillaume Carlier$^{\flat }$}
\author[V. Chernozhukov]{Victor Chernozhukov$^{\lozenge }$}
\author[G. De Bie]{Gwendoline De Bie{$^{\mathsection}$}}
\author[A. Galichon]{Alfred Galichon$^{\dag }$ }
\date{7/31/2019 \ (First draft). This draft: 7/25/2020.\\
\indent This is a pre-print of an article published in \textit{Empirical Economics} (2020). The final authenticated version is available online at: \url{https://doi.org/10.1007/s00181-020-01919-y}.\\
{\indent$^{\flat }$CEREMADE, UMR CNRS 7534, PSL, Universit\'{e} }Paris IX
Dauphine, Pl. de Lattre de Tassigny, 75775 Paris Cedex 16, FRANCE, and
MOKAPLAN Inria Paris; carlier@ceremade.dauphine.fr.{\ }\\
{\indent$^{\lozenge }$}Department of Economics, MIT, 50 Memorial Drive,
E52-361B, Cambridge, MA 02142, USA; vchern@mit.edu.\\
{\indent$^{\mathsection}$}DMA, ENS Paris; gwendoline.de.bie@ens.fr. Funding from R\'{e}gion
Ile-de-France grant is acknowledged.\\
{\indent$^{\dag }$}Economics and Mathematics Departments, New York
University, 70 Washington Square South, New York, NY 10013, USA;
ag133@nyu.edu. Funding from NSF grant DMS-1716489 is acknowledged}

\begin{abstract}
In this paper, we first revisit the Koenker and Bassett variational approach
to (univariate) quantile regression, emphasizing its link with latent factor
representations and correlation maximization problems. We then review the
multivariate extension due to Carlier et al. (2016, 2017) which relates
vector quantile regression to an optimal transport problem with mean
independence constraints. We introduce an entropic regularization of this
problem, implement a gradient descent numerical method and illustrate its
feasibility on univariate and bivariate examples. 

\vspace{0.1cm} \emph{Keywords:} vector quantile regression, optimal
transport with mean independence constraints, latent factors, entropic
regularization

\vspace{0.1cm} \emph{JEL Classification:} C51, C60

\vspace{0.5cm}
\end{abstract}

\maketitle

\pagebreak 

\section{Introduction}

Quantile regression, introduced \ by Koenker and Bassett (1978), has become
a very popular tool for analyzing the response of the whole distribution of
a dependent variable to a set of predictors. It is a far-reaching
generalization of the median regression, allowing for a predition of any
quantile of the distribution. We briefly recall classical quantile
regression. For $t\in \left[ 0,1\right] $, it is well-known that the $t$%
-quantile of $\varepsilon =Y-q_{t}\left( x\right) $ given $X=x$ minimizes
the loss function $\mathbb{E}\left[ t\varepsilon ^{+}+\left( 1-t\right)
\varepsilon ^{-}|X\right] $, or equivalently $\mathbb{E}\left[ \varepsilon
^{+}+\left( t-1\right) \varepsilon |X\right] $. As a result, if $q_{t}\left(
x\right) $ is specified under the parametric form $q_{t}\left( x\right)
=\beta_{t}^{\top } x+\alpha _{t}$, it is natural to estimate $\alpha _{t}$
and $\beta _{t}$ by minimizing the loss 
\begin{equation*}
\min_{\alpha ,\beta }\mathbb{E}\left[ \left( Y-\beta^{\top} X -\alpha
\right) ^{+}+\left( 1-t\right) \left( \beta^{\top }X +\alpha \right) \right]
.
\end{equation*}

While the previous optimization problem estimates $\alpha _{t}$ and $\beta
_{t}$ for pointwise values of $t$, if one would like to estimate the whole
curve $t\mapsto \left( \alpha _{t},\beta _{t}\right) $, one simply should
construct the loss function by integrating the previous loss functions over $%
t\in \left[ 0,1\right] $, and thus the curve $t\mapsto \left( \alpha
_{t},\beta _{t}\right) $ minimizes%
\begin{equation*}
\min_{\left( \alpha _{t},\beta _{t}\right) _{t\in \left[ 0,1\right]
}}\int_{0}^{1}\mathbb{E}\left[ \left( Y-\beta_t^{\top} X-\alpha _{t}\right)
^{+}+\left( 1-t\right) \left( \beta_t^{\top} X+\alpha _{t}\right) \right] dt.
\end{equation*}

As it is known since the original work by Koenker and Bassett, this problem
has an (infinite-dimensional) linear programming formulation. Defining $%
P_{t}=\left( Y-\beta_t^{\top} X-\alpha _{t}\right) ^{+}$ as the positive
deviations of $Y$ with respect to their predicted quantiles $\beta_t^{\top}
X+\alpha _{t}$, we have $P_{t}\geq 0$ and $\left( Y-\beta_t^{\top} X-\alpha
_{t}\right) ^{-}=P_{t}-Y+\beta_t^{\top} X+\alpha _{t}\geq 0$, so the problem
reformulates as\footnote{%
Whenever we write a variable in brackets after a constraint, as $[V_t]$ in (%
\ref{dual_intro}), we mean that this variable plays the role of a multiplier.%
} 
\begin{align}
\min_{P_{t}\geq 0,\beta _{t},\alpha _{t}}& \int_{0}^{1}\mathbb{E}\left[
P_{t}+\left( 1-t\right) \left( \beta_t^{\top} X+\alpha _{t}\right) \right] dt
\label{dual_intro} \\
s.t.~& P_{t}-Y+\beta_t^{\top} X+\alpha _{t}\geq 0~\left[ V_{t}\right]  \notag
\end{align}%
which we will call \textquotedblleft dual formulation\textquotedblright\ of
the classical quantile regression problem\footnote{%
It may seem awkward to start with the \textquotedblleft
dual\textquotedblright\ formulation before giving out the \textquotedblleft
primal\textquotedblright\ one, and the \textquotedblleft
primal\textquotedblright\ being the dual to the \textquotedblleft
dual,\textquotedblright\ this choice of labeling is pretty arbitrary.
However, our choice is motivated by consistency with optimal transport
theory, introduced below.}. To the dual formulation corresponds a primal one
(dual to the dual), which is formally obtained by a minimax formulation%
\begin{equation*}
\min_{P_{t}\geq 0,\beta _{t},\alpha _{t}}\max_{V_{t}\geq 0}\int_{0}^{1}%
\mathbb{E}\left[ P_{t}+\left( 1-t\right) \left( \beta_t^{\top} X+\alpha
_{t}\right) +V_{t}Y-V_{t}P-V_{t}\beta_t^{\top} X-V_{t}\alpha _{t}\right] dt
\end{equation*}%
thus%
\begin{equation*}
\max_{V_{t}\geq 0}\int_{0}^{1}\mathbb{E}\left[ V_{t}Y\right]
dt+\min_{P_{t}\geq 0,\beta _{t},\alpha _{t}}\int_{0}^{1}\mathbb{E}\left[
\left( 1-V_{t}\right) P_{t}+\beta _{t}^{\top }\left( \left( 1-t-V_{t}\right)
X\right) +V_{t}\left( Y-\alpha _{t}\right) \right] dt
\end{equation*}%
hence we arrive at the primal formulation 
\begin{align}
& \max_{V_{t}\geq 0}\int_{0}^{1}\mathbb{E}\left[ YV_{t}\right] dt
\label{primal_intro} \\
s.t.~& V_{t}\leq 1~\left[ P_{t}\geq 0\right]  \notag \\
& \mathbb{E}\left[ V_{t}X\right] =\left( 1-t\right) \mathbb{E}\left[ X\right]
~\left[ \beta _{t}\right]  \notag \\
& \mathbb{E}\left[ V_{t}\right] =\left( 1-t\right) ~\left[ \alpha _{t}\right]
\notag
\end{align}

If $V_{t}$ and $\left( \alpha _{t},\beta _{t}\right) $ are solutions to the
above primal and dual programs, complementary slackness yields $\mathbf{1}%
_{\left\{ Y>\beta_t^{\top} X+\alpha _{t}\right\}} \leq V_{t}\leq \mathbf{1}%
_{\left\{ Y\geq \beta_t^{\top} X+\alpha _{t}\right\}} $, hence if $\left(
X,Y\right) $ has a continuous distribution, then for any $\left( \alpha
,\beta \right) $, ${\mathbb{P}}\left( Y-\beta^{\top} X -\alpha =0\right) =0$%
, and therefore one has almost surely%
\begin{equation*}
V_{t}=\mathbf{1}_{\left\{ Y\geq \beta_t^{\top} X+\alpha _{t}\right\}}.
\end{equation*}

Koenker and Ng (2005) impose a monotonicity constraint of the estimated
quantile curves. Indeed, if $\beta_t^{\top} x+\alpha _{t}$ is the $t$%
-quantile of the conditional distribution of $Y$ given $X=x$, the curve $%
t\mapsto \beta_t^{\top} x+\alpha _{t}$ should be nondecreasing. Hence, these
authors impose a natural constraint on the dual, that is $\beta_t^{\top} X +
\alpha_{t}\geq \beta_{t^{\prime }}^{\top} X+\alpha_{t^{\prime}}$ for $t\geq
t^{\prime } $, and they incorporate this constraint into~(\ref{dual_intro}),
yielding%
\begin{align*}
\min_{P_{t}\geq 0,N_{t}\geq 0,\beta _{t},\alpha _{t}}& \int_{0}^{1}\mathbb{E}%
\left[ P_{t}+\left( 1-t\right) \left( \beta_t^{\top} X+\alpha _{t}\right) %
\right] dt \\
s.t.~& P_{t}-N_{t}=Y-\beta_t^{\top} X-\alpha _{t}~\left[ V_{t}\right] \\
& t\geq t^{\prime } \Rightarrow \beta_t^{\top} X+\alpha_{t}\geq
\beta_{t^{\prime}}^{\top} X +\alpha_{t^{\prime}}.
\end{align*}
Note that if $t\mapsto \beta_t^{\top} x+\alpha _{t}$ is nondecreasing, then $%
\mathbf{1}_{\left\{ y\geq \beta_t^{\top} x+\alpha _{t}\right\}}$ should be
nonincreasing. Therefore, in that case, $V_{t}$ should be nonincreasing in $t
$, which allows us to impose a monotonicity constraint on the primal
variable $V_{t}$ instead of a monotonicity constraint on the dual variables $%
\beta _{t}$ and $\alpha_t$. This is precisely the problem we look at.
Consider 
\begin{align}
& \max_{V_{t}}\int_{0}^{1}\mathbb{E}\left[ YV_{t}\right] dt
\label{intro_primalConstrainted} \\
s.t.~& V_{t}\geq 0~\left[ N_{t}\geq 0\right]  \notag \\
& V_{t}\leq 1~\left[ P_{t}\geq 0\right]  \notag \\
& \mathbb{E}\left[ V_{t}X\right] =\left( 1-t\right) \mathbb{E}\left[ X\right]
~\left[ \beta _{t}\right]  \notag \\
& \mathbb{E}\left[ V_{t}\right] =\left( 1-t\right) ~\left[ \alpha _{t}\right]
\notag \\
& t\geq t^{\prime } \Rightarrow V_t \leq V_{t^{\prime }}.  \notag
\end{align}

Let us now take a look at a sample version of this problem. Here, we observe
as sample $\left( X_{i},Y_{i}\right) $ for $i\in \left\{ 1,...,N\right\} $.
We shall discretize the probability space $\left[ 0,1\right] $ into $T$
points, $t_{1}=0<t_{2}<...<t_{T}\leq 1$. Let $\bar{x}$ be the $1\times K$
row vector whose $k$-th entry is $\sum_{1\leq i\leq n}X_{ik}/N$. The sample
analog of~(\ref{intro_primalConstrainted}) is%
\begin{align*}
& \max_{V_{\tau i}\geq 0}\sum_{\substack{ 1\leq i\leq N  \\ 1\leq \tau \leq
T }}V_{\tau i}Y_{i} \\
& V_{\tau i}\leq 1 \\
& \frac{1}{N}\sum_{1\leq i\leq N}V_{\tau i}X_{ik}=\left( 1-t_{\tau }\right) 
\bar{x}_{k} \\
& \frac{1}{N}\sum_{1\leq i\leq N}V_{\tau i}=\left( 1-t_{\tau }\right) \\
& V_{\tau 1}\geq V_{\tau 2}\geq ...\geq V_{\tau \left( N-1\right) }\geq
V_{\tau ,N}\geq 0.
\end{align*}

Denoting $\mathbf{t}$ the $T\times 1$ row matrix with entries $t_{\tau }$,
and $D$ a $T\times T$ matrix with ones on the main diagonal, and $-1$ on the
diagonal just below the main diagonal, and $0$ elsewhere, the condition $%
V_{\tau 1}\geq V_{\tau 2}\geq ...\geq V_{\tau \left( N-1\right) }\geq
V_{\tau ,N}\geq 0$ reexpresses as $V^{\top }D\geq 0$, and the program
rewrites%
\begin{align*}
& \max_{V}1_{N}^{\top }VY \\
& \frac{1}{N}VX=\left( 1_{T}-\mathbf{t}\right) \bar{x} \\
& \frac{1}{N}V1_{N}=\left( 1_{T}-\mathbf{t}\right) \\
& V^{\top }D1_{T}=1_{N} \\
& V^{\top }D\geq 0.
\end{align*}

Setting $\pi =D^{\top }V/N$, and $U=D^{-1}1_{N}=\left( 1/T,2/T,...,1\right) $%
, $\mu =D^{\top }\left( 1_{T}-\mathbf{t}\right) =\left( 1/T,...,1/T\right) $%
, and $\nu=1_{N}/N$, one can reformulate the problem as%
\begin{align*}
& \max_{\pi \geq 0}\sum_{\substack{ 1\leq \tau \leq T  \\ 1\leq i\leq N}}\pi
_{\tau i}U_{\tau }Y_{i} \\
& \sum_{1\leq \tau \leq T}\pi _{\tau i}=\nu_{i} \\
& \sum_{i=1}^{N}\pi _{\tau i}=\mu _{\tau }\text{ } \\
& \sum_{1\leq i\leq N}\pi _{\tau i}X_{ik}=\mu _{\tau }\bar{x}_{k}
\end{align*}%
which rewrites in the population as 
\begin{align}
\max_{\left( U,X,Y\right) \sim \pi }& \mathbb{E}_{\pi }\left[ UY\right]
\label{VQRpb} \\
s.t. \; & U\sim \mathcal{U}\left( \left[ 0,1\right] \right)  \notag \\
& \left( X,Y\right) \sim \nu  \notag \\
& \mathbb{E}\left[ X|U\right] =\mathbb{E}\left[ X\right] .  \notag
\end{align}

Note that this is a direct extension of the Monge-Kantorovich problem of
optimal transport -- in fact it boils down to it when the last constraint is
absent. This should not be surprising, given the connection between optimal
transport, as recalled below. In the present paper, we introduce the \emph{%
Regularized Vector Quantile Regression} (RVQR) problem, which consists of
adding an entropic regularization term in the expression~(\ref{VQRpb}),
which yields, for a given data distribution $\nu$,%
\begin{align}
\max_{\left( U,X,Y\right) \sim \pi }& \mathbb{E}_{\pi }\left[ UY\right]
-\varepsilon \mathbb{E}_{\pi }\left[ \ln \pi \left( U,X,Y\right) \right]
\label{RVQRpb} \\
s.t. \; & U\sim \mathcal{U}\left( \left[ 0,1\right] \right)  \notag \\
& \left( X,Y\right) \sim \nu  \notag \\
& \mathbb{E}\left[ X|U\right] =\mathbb{E}\left[ X\right] .  \notag
\end{align}%
Due to smoothness and regularity, the regularized problem~(\ref{RVQRpb})
enjoys computational and analytical properties that are missing from the
original problem~(\ref{VQRpb}). In particular, the dual to~(\ref{VQRpb}) is
a smooth, unconstrained problem that can be solved by computational methods.
While here, unlike in the context of stadard optimal transport, the
Kullback-Leibler divergence projection onto the mean-independence constraint
is not in closed form, we can use Nesterov's gradient descent acceleration,
which gives optimal convergence rates for first-order methods.

The present paper in part provides a survey of previous results, and in part
conveys new results. In the vein of the previous papers on the topic (2016,
2017), this paper seeks to apply the optimal transport toolbox to quantile
regression. In contrast with these papers, a particular focus in the present
paper is to propose a regularized version of the problem as well as new
computational methods. The two main new contributions of the paper are (1) a
connection with shape-constrained classical regression (section 4), and (2)
the introduction of the regularized vector quantile regression problem
(RVQR) along with a duality theorem for that problem (section 6).

The paper is organized as follows. Section 2 will offer reminders on the
notion of quantile; section 3 will review the previous results of Carlier et
al. (2016, 2017) on the \textquotedblleft specified\textquotedblright\ case;
section 4 offers a new result (theorem 4.1) on the comparison with the
shape-constrained classical quantile regression; and section 5 will review
results on the multivariate case. Section 6 introduces RVQR\ and introduces
results relevant for that problem, in particular a duality result in that
case (theorem 6.1).

\section{Several characterizations of quantiles}

\label{sec-char-quant}

Throughout the paper, $(\Omega, {\mathcal{F}}, \mathbb{P})$ will be some
fixed nonatomic space\footnote{%
One way to define the nonatomicity of $(\Omega, {\mathcal{F}}, \mathbb{P})$
is by the existence of a uniformly distributed random variable on this
space, this somehow ensures that the space is rich enough so that there
exists random variables with prescribed law. If, on the contrary, the space
is finite for instance only finitely supported probability measures can be
realized as the law of such random variables.} probability. Given a random
vector $Z$ with values in ${\mathbb{R}}^k$ defined on this space we will
denote by ${\mathscr{Law}}(Z)$ the law of $Z$, given a probability measure $%
\theta$ on ${\mathbb{R}}^k$, we shall often write $Z\sim \theta$ to express
that ${\mathscr{Law}}(Z)=\theta$. Independence of two random variables $Z_1$
and $Z_2$ will be denoted as $Z_1 \perp \! \! \! \perp Z_2$.

\subsection{Quantiles}

\label{subsec-quant}

Let $Y$ be some univariate random variable defined on $(\Omega, {\mathcal{F}}%
, \mathbb{P})$. Denoting by $F_Y$ the distribution function of $Y$: 
\begin{equation*}
F_Y(\alpha):=\mathbb{P}(Y\le \alpha), \; \forall \alpha\in {\mathbb{R}}
\end{equation*}
the \emph{quantile} function of $Y$, $Q_Y=F_Y^{-1}$ is the generalized
inverse of $F_Y$ given by the formula: 
\begin{equation}  \label{defqy}
Q_Y(t):=\inf\{\alpha \in {\mathbb{R}} \; : \; F_Y(\alpha)>t\} 
\mbox{ for all
} t\in (0,1).
\end{equation}
Let us now recall two well-known facts about quantiles:

\begin{itemize}
\item $\alpha=Q_Y(t)$ is a solution of the convex minimization problem 
\begin{equation}  \label{minquant}
\min_{\alpha} \{{\mathbb{E}}((Y-\alpha)^+)+ \alpha(1-t)\}
\end{equation}

\item there exists a uniformly distributed random variable $U$ such that $%
Y=Q_Y(U)$. Moreover, among uniformly distributed random variables, $U$ is
maximally correlated\footnote{%
In fact for (\ref{oteasy}) to make sense one needs some integrabilty of $Y$
i.e. ${\mathbb{E}}(\vert Y\vert) <+\infty$.} to $Y$ in the sense that it
solves 
\begin{equation}  \label{oteasy}
\max \{ {\mathbb{E}}(VY), \; V\sim \mu\}
\end{equation}
where $\mu:={\mathcal{U}}([0,1])$ is the uniform measure on $[0,1]$.

Of course, when ${\mathscr{Law}}(Y)$ has no atom, i.e. when $F_{Y}$ is
continuous, $U$ is unique and given by $U=F_{Y}(Y)$. Problem (\ref{oteasy})
is the easiest example of optimal transport problem one can think of. The
decomposition of a random variable $Y$ as the composed of a monotone
nondecreasing function and a uniformly distributed random variable is called
a \emph{polar factorization} of $Y$. The existence of such decompositions
goes back to Ryff (1970) and the extension to the multivariate case (by
optimal transport) is due to Brenier (1991).
\end{itemize}

We therefore see that there are basically two different approaches to study
or estimate quantiles:

\begin{itemize}
\item the \emph{local} or "$t$ by $t$" approach which consists, for a fixed
probability level $t$, in using directly formula (\ref{defqy}) or the
minimization problem (\ref{minquant}) (or some approximation of it), this
can be done very efficiently in practice but has the disadvantage of
forgetting the fundamental global property of the quantile function: it
should be monotone in $t$,

\item the global approach (or polar factorization approach), where quantiles
of $Y$ are defined as all nondecreasing functions $Q$ for which one can
write $Y=Q(U)$ with $U$ uniformly distributed. In this approach, one rather
tries to recover directly the whole monotone function $Q$ (or the uniform
variable $U$ that is maximally correlated to $Y$). Therefore this is a
global approach for which one should rather use the optimal transport
problem (\ref{oteasy}).
\end{itemize}

\subsection{Conditional quantiles}

\label{subsec-cond-quant}

Let us assume now that, in addition to the random variable $Y$, we are also
given a random vector $X\in {\mathbb{R}}^N$ which we may think of as being a
list of explanatory variables for $Y$. We are primarily interested in the
dependence between $Y$ and $X$ and in particular the conditional quantiles
of $Y$ given $X=x$. Let us denote by $\nu$ the joint law of $(X,Y)$ by $\nu$
the law of $X$, and by $\nu(. \vert x)$ the conditional law of $Y$ given $%
X=x $: 
\begin{equation}  \label{notationslawxy}
\nu:={\mathscr{Law}}(X,Y), \; m:={\mathscr{Law}}(X), \; \nu(.\vert x):={%
\mathscr{Law}}(Y \vert X=x)
\end{equation}
which in particular yields 
\begin{equation*}
\mbox{d} \nu(x,y) = \mbox{d}\nu(y\vert x) \mbox{d} m(x).
\end{equation*}

\smallskip

We then denote by $F(x,y)=F_{Y\vert X=x}(y)$ the conditional cdf: 
\begin{equation*}
F(x,y):=\mathbb{P}(Y\le y \vert X=x)
\end{equation*}
and $Q(x,t)$ the conditional quantile 
\begin{equation*}
Q(x,t):=\inf\{\alpha\in {\mathbb{R}} \; : \; F(x,\alpha)>t\}, \; \forall
t\in (0,1).
\end{equation*}
For the sake of simplicity, we shall assume that for $m={\mathscr{Law}}(X)$%
-almost every $x\in {\mathbb{R}}^N$ ($m$-a.e. $x$ for short), one has 
\begin{equation}  \label{hypo1}
t\mapsto Q(x,t) \mbox{ is
continuous and increasing}
\end{equation}
so that for $m$-a.e. $x$, $F(x, Q(x,t))=t$ for every $t\in (0,1)$ and $Q(x,
F(x,y))=y$ for every $y$ in the support of $\nu(.\vert x)$.

Let us now define the random variable 
\begin{equation}  \label{defdeuu}
U:=F(X,Y),
\end{equation}
then by construction: 
\begin{equation*}
\begin{split}
\mathbb{P}(U< t\vert X=x)&=\mathbb{P}(F(x,Y)<t \vert X=x)=\mathbb{P}%
(Y<Q(x,t) \vert X=x) \\
&=F(x,Q(x,t))=t.
\end{split}%
\end{equation*}
We deduce that $U$ is uniformly distributed and independent from $X$ (since
its conditional cdf does not depend on $x$). Moreover since $U=F(X,Y)=F(X,
Q(X,U))$ it follows from (\ref{hypo1}) that one has the representation 
\begin{equation*}
Y=Q(X,U)
\end{equation*}
in which $U$ can naturally be interpreted as a latent factor.

This easy remark leads to a conditional polar factorization of $Y$ through
the pointwise relation $Y=Q(X,U)$ with $Q(X,.)$ nondecreasing and $U\sim \mu$%
, $U\perp \! \! \! \perp X$. We would like to emphasize now that there is a
variational principle behind this conditional decomposition. Let us indeed
consider the variant of the optimal transport problem (\ref{oteasy}) where
one further requires $U$ to be independent from the vector of regressors $X$%
: 
\begin{equation}  \label{otindep1d}
\max \{ {\mathbb{E}}(VY), \; {\mathscr{Law}}(V)=\mu, \; V \perp \! \! \!
\perp X \}.
\end{equation}
then we have

\begin{prop}
\label{univaropt} If ${\mathbb{E}}(\vert Y\vert)<+\infty$ and (\ref{hypo1})
holds, the random variable $U$ defined in (\ref{defdeuu}) solves (\ref%
{otindep1d}).
\end{prop}

\begin{proof}
Let $V$ be admissible for (\ref{otindep1d}). Let us define for $x\in %
\mathop{\mathrm{spt}}\nolimits(m)$ and $t\in [0,1]$, 
\begin{equation*}
\varphi(x,t):=\int_0^t Q(x,s) \mbox{d}s.
\end{equation*}
We first claim that $\varphi(X,U)$ is integrable, indeed we obviously have 
\begin{equation*}
\vert \varphi(X,U)\vert \leq \int_0^1 \vert Q(X,s)\vert \mbox{d}s
\end{equation*}
hence 
\begin{equation*}
\begin{split}
{\mathbb{E}}(\vert \varphi(X,U)\vert) &\leq \int_{{\mathbb{R}}^N} \int_0^1
\vert Q(x,s)\vert \mbox{d} \mu(s)\; \mbox{d} m(x) \\
&=\int_{{\mathbb{R}}^N} \int_{\mathbb{R}} \vert y \vert \mbox{d} \nu(y\vert
x) \mbox{d} m(x) ={\mathbb{E}}(\vert Y \vert)<+\infty
\end{split}%
\end{equation*}
where we have used in the second line the fact that the image of $\mu$ by $%
Q(x,.)$ is $\nu(.\vert x)$. Since $\varphi(x,.)$ is convex and $Y=\frac{%
\partial \; \varphi} {\partial u} (X,U)$ the pointwise inequality 
\begin{equation*}
\varphi(X,V)-\varphi(X,U)\geq Y (V-U)
\end{equation*}
holds almost surely. But since ${\mathscr{Law}}(X,V)={\mathscr{Law}}(X,U)$
integrating the previous inequality yields 
\begin{equation*}
{\mathbb{E}}(\varphi(X,V)-\varphi(X,U))=0 \geq {\mathbb{E}}(Y (V-U)).
\end{equation*}
\end{proof}

\section{Specified and quasi-specified quantile regression}

\subsection{Specified quantile regression}

Since the seminal work of Koenker and Bassett (1978), it has been widely
accepted that a convenient way to estimate conditional quantiles is to
stipulate an affine form with respect to $x$ for the conditional quantile.
Since a quantile function should be monotone in its second argument, this
leads to the following definition

\begin{defi}
\label{defispec} Quantile regression is specified if there exist $(\alpha,
\beta)\in C([0,1], {\mathbb{R}})\times C([0,1], {\mathbb{R}}^N)$ such that
for $m$-a.e. $x$ 
\begin{equation}  \label{monqr}
t\mapsto \alpha(t)+\beta(t)^{\top} x \mbox{ is increasing on $[0,1]$}
\end{equation}
and 
\begin{equation}  \label{linearcq}
Q(x,t)=\alpha(t)+ \beta(t)^{\top} x,
\end{equation}
for $m$-a.e. $x$ and every $t\in [0,1]$. If (\ref{monqr})-(\ref{linearcq})
hold, quantile regression is specified with regression coefficients $%
(\alpha, \beta)$.
\end{defi}

Specification of quantile regression can be characterized by the validity of
an affine in $X$ representation of $Y$ with a latent factor:

\begin{prop}
Let $(\alpha, \beta)$ be continuous and satisfy (\ref{monqr}). Quantile
regression is specified with regression coefficients $(\alpha, \beta)$ if
and only if there exists $U$ such that 
\begin{equation}  \label{polarfqrind}
Y=\alpha(U)+ \beta(U)^{\top} X \mbox{ almost surely} , \; {\mathscr{Law}}%
(U)=\mu, \; U \perp \! \! \! \perp X.
\end{equation}
\end{prop}

\begin{proof}
The fact that specification of quantile regression implies decomposition (%
\ref{polarfqrind}) has already been explained in paragraph \ref%
{subsec-cond-quant}. Let us assume (\ref{polarfqrind}), and compute 
\begin{equation*}
\begin{split}
F(x, \alpha(t)+\beta(t)^{\top} x)&=\mathbb{P}(Y\leq
\alpha(t)+\beta(t)^{\top} x\vert X=x) \\
&= \mathbb{P}(\alpha(U)+ \beta(U)^{\top} x \leq \alpha(t)+\beta(t)^{\top}
x\vert X=x) \\
&=\mathbb{P}(U\leq t \vert X=x)=\mathbb{P}(U\le t)=t
\end{split}%
\end{equation*}
so that $Q(x,t)=\alpha(t)+\beta(t)^{\top} x$.
\end{proof}

\subsection{Quasi-specified quantile regression}

Let us now assume that both $X$ and $Y$ are integrable 
\begin{equation}  \label{finitemoment}
{\mathbb{E}}(\Vert X\Vert + \vert Y\vert)<+\infty
\end{equation}
and normalize, without loss of generality, $X$ in such a way that 
\begin{equation}  \label{normalizX}
{\mathbb{E}}(X)=0.
\end{equation}

Koenker and Bassett showed that, for a fixed probability level $t$, the
regression coefficients $(\alpha ,\beta )$ can be estimated by quantile
regression i.e. the minimization problem 
\begin{equation}
\inf_{(\alpha ,\beta )\in {\mathbb{R}}^{1+N}}{\mathbb{E}}(\rho _{t}(Y-\alpha
-\beta^{\top}X))  \label{kb0}
\end{equation}%
where the penalty $\rho _{t}$ is given by $\rho _{t}(z):=tz^-+(1-t)z^+$ with 
$z^-$ and $z^+$ denoting the negative and positive parts of $z$. For further
use, note that (\ref{kb0}) can be conveniently be rewritten as 
\begin{equation}
\inf_{(\alpha ,\beta )\in {\mathbb{R}}^{1+N}}\{{\mathbb{E}}((Y-\alpha
-\beta^{\top}X)^+)+(1-t)\alpha \}.  \label{kb1}
\end{equation}%
As noticed by Koenker and Bassett, this convex program admits as dual
formulation 
\begin{equation}
\sup \{{\mathbb{E}}(V_{t}Y))\;:\;V_{t}\in \lbrack 0,1],\;{\mathbb{E}}%
(V_{t})=(1-t),\;{\mathbb{E}}(V_{t}X)=0\}.  \label{dt}
\end{equation}%
An optimal $(\alpha ,\beta )$ for (\ref{kb1}) and an optimal $V_{t}$ in (\ref%
{dt}) are related by the complementary slackness condition: 
\begin{equation}
Y>\alpha +\beta^{\top}X\Rightarrow V_{t}=1,\mbox{ and }\;Y<\alpha
+\beta^{\top}X\Rightarrow V_{t}=0.
\end{equation}%
Note that $\alpha $ appears naturally as a Lagrange multiplier associated to
the constraint ${\mathbb{E}}(V_{t})=(1-t)$ and $\beta $ as a Lagrange
multiplier associated to ${\mathbb{E}}(V_{t}X)=0$.

To avoid mixing i.e. the possibility that $V_t$ takes values in $(0,1)$, it
will be convenient to assume that $\nu={\mathscr{Law}}(X,Y)$ gives zero mass
to nonvertical hyperplanes i.e. 
\begin{equation}  \label{nonvert}
\mathbb{P}(Y=\alpha+\beta^{\top} X)=0, \; \forall (\alpha, \beta)\in {%
\mathbb{R}}^{1+N}.
\end{equation}
We shall also consider a nondegeneracy condition on the (centered) random
vector $X$ which says that its law is not supported by any hyperplane%
\footnote{%
if ${\mathbb{E}}(\Vert X\Vert^2)<+\infty$ then (\ref{nondegX}) amounts to
the standard requirement that ${\mathbb{E}}(X X^{\top})$ is nonsingular.}: 
\begin{equation}  \label{nondegX}
\; \mathbb{P}(\beta^{\top} X=0)<1, \; \forall \beta\in {\mathbb{R}}%
^N\setminus\{0\}.
\end{equation}

Thanks to (\ref{nonvert}), we may simply write 
\begin{equation}  \label{frombetatoU}
V_t=\mathbf{1}_{\{Y>\alpha +\beta^{\top} X\}}
\end{equation}
and thus the constraints ${\mathbb{E}}(V_t)=(1-t)$, ${\mathbb{E}}(XV_t)=0$
read 
\begin{equation}  \label{normaleq}
{\mathbb{E}}( \mathbf{1}_{\{Y> \alpha+ \beta^{\top} X \}})=\mathbb{P}(Y>
\alpha+ \beta^{\top} X) = (1-t),\; {\mathbb{E}}(X \mathbf{1}_{\{Y> \alpha+
\beta^{\top} X \}} ) =0
\end{equation}
which simply are the first-order conditions for (\ref{kb1}).

Any pair $(\alpha, \beta)$ which solves\footnote{%
Uniqueness will be discussed later on.} the optimality conditions (\ref%
{normaleq}) for the Koenker and Bassett approach will be denoted 
\begin{equation*}
\alpha=\alpha^{QR}(t), \beta=\beta^{QR}(t)
\end{equation*}
and the variable $V_t$ solving (\ref{dt}) given by (\ref{frombetatoU}) will
similarly be denoted $V_t^{QR}$ 
\begin{equation}  \label{utqr}
V_t^{QR}:=\mathbf{1}_{\{Y>\alpha^{QR}(t) +\beta^{QR}(t)^{\top} X\}}.
\end{equation}

Note that in the previous considerations the probability level $t$ is fixed,
this is what we called the "$t$ by $t$" approach. For this approach to be
consistent with conditional quantile estimation, if we allow $t$ to vary we
should add an additional monotonicity requirement:

\begin{defi}
Quantile regression is quasi-specified\footnote{%
If quantile regression is specified and the pair of functions $(\alpha,
\beta)$ is as in definition \ref{defispec}, then for every $t$, $(\alpha(t),
\beta(t))$ solves the conditions (\ref{normaleq}). This shows that
specification implies quasi-specification.} if there exists for each $t$, a
solution $(\alpha^{QR}(t), \beta^{QR}(t))$ of (\ref{normaleq}) (equivalently
the minimization problem (\ref{kb0})) such that $t\in [0,1]\mapsto
(\alpha^{QR}(t), \beta^{QR}(t))$ is continuous and, for $m$-a.e. $x$ 
\begin{equation}  \label{monqrqs}
t\mapsto \alpha^{QR}(t)+\beta^{QR}(t)^{\top} x \mbox{ is increasing on
$[0,1]$}.
\end{equation}
\end{defi}

A first consequence of quasi-specification is given by

\begin{prop}
\label{qrqsdec} Assume (\ref{hypo1})-(\ref{finitemoment})-(\ref{normalizX})
and (\ref{nonvert}). If quantile regression is quasi-specified and if we
define $U^{QR}:=\int_0^1 V_t^{QR} dt$ (recall that $V_t^{QR}$ is given by (%
\ref{utqr})) then:

\begin{itemize}
\item $U^{QR}$ is uniformly distributed,

\item $X$ is mean-independent from $U^{QR}$ i.e. ${\mathbb{E}}(X\vert
U^{QR})={\mathbb{E}}(X)=0$,

\item $Y=\alpha^{QR}(U^{QR})+ {\beta^{QR}}(U^{QR})^{\top} X$ almost surely.
\end{itemize}

Moreover $U^{QR}$ solves the correlation maximization problem with a
mean-independence constraint: 
\begin{equation}  \label{maxcorrmi}
\max \{ {\mathbb{E}}(VY), \; {\mathscr{Law}}(V)=\mu, \; {\mathbb{E}}(X\vert
V)=0\}.
\end{equation}
\end{prop}

\begin{proof}
Obviously 
\begin{equation*}
V_t^{QR}=1\Rightarrow U^{QR} \ge t, \mbox{ and } \; U^{QR}>t \Rightarrow
V_t^{QR}=1
\end{equation*}
hence $\mathbb{P}(U^{QR}\ge t)\ge \mathbb{P}(V_t^{QR}=1)=\mathbb{P}(Y>
\alpha^{QR}(t)+\beta^{QR}(t)^{\top} X)=(1-t)$ and $\mathbb{P}(U^{QR}> t)\le 
\mathbb{P}(V_t^{QR}=1)=(1-t)$ which proves that $U^{QR}$ is uniformly
distributed and $\{U^{QR}>t\}$ coincides with $\{V^{QR}_t=1\}$ up to a set
of null probability. We thus have ${\mathbb{E}}(X \mathbf{1}_{U^{QR}>t})={%
\mathbb{E}}(X V_t^{QR})=0$, by a standard approximation argument we deduce
that ${\mathbb{E}}(Xf(U^{QR}))=0$ for every $f\in C([0,1], {\mathbb{R}})$
which means that $X$ is mean-independent from $U^{QR}$.

As already observed $U^{QR}>t$ implies that $Y>\alpha^{QR}(t)+%
\beta^{QR}(t)^{\top} X$ in particular $Y\ge
\alpha^{QR}(U^{QR}-\delta)+\beta^{QR}(U^{QR}- \delta)^{\top} X$ for $\delta>0
$, letting $\delta\to 0^+$ and using the continuity of $(\alpha^{QR},
\beta^{QR})$ we get $Y\ge \alpha^{QR}(U^{QR})+\beta^{QR}(U^{QR})^{\top} X$.
The converse inequality is obtained similarly by remarking that $U^{QR}<t$
implies that $Y\le \alpha^{QR}(t)+\beta^{QR}(t)^{\top} X$.

Let us now prove that $U^{QR}$ solves (\ref{maxcorrmi}). Take $V$ uniformly
distributed, such that $X$ is mean-independent from $V$ and set $V_t:=%
\mathbf{1}_{\{V>t \}}$, we then have ${\mathbb{E}}(X V_t)=0$, ${\mathbb{E}}%
(V_t)=(1-t)$ but since $V_t^{QR}$ solves (\ref{dt}) we have ${\mathbb{E}}%
(V_t Y)\le {\mathbb{E}}(V_t^{QR}Y)$. Observing that $V=\int_0^1 V_t dt$ and
integrating the previous inequality with respect to $t$ gives ${\mathbb{E}}%
(VY)\le {\mathbb{E}}(U^{QR}Y)$ so that $U^{QR}$ solves (\ref{maxcorrmi}).
\end{proof}

Let us continue with a uniqueness argument for the mean-independent
decomposition given in proposition \ref{qrqsdec}:

\begin{prop}
\label{uniquedec} Assume (\ref{hypo1})-(\ref{finitemoment})-(\ref{normalizX}%
)-(\ref{nonvert}) and (\ref{nondegX}). Let us assume that 
\begin{equation*}
Y=\alpha(U)+\beta(U)^{\top} X=\overline{\alpha} (\overline{U})+ \overline{%
\beta}(\overline{U})^{\top} X
\end{equation*}
with:

\begin{itemize}
\item both $U$ and $\overline{U}$ uniformly distributed,

\item $X$ is mean-independent from $U$ and $\overline{U}$: ${\mathbb{E}}%
(X\vert U)={\mathbb{E}}(X\vert \overline{U})=0$,

\item $\alpha, \beta, \overline{\alpha}, \overline{\beta}$ are continuous on 
$[0,1]$,

\item $(\alpha, \beta)$ and $(\overline{\alpha}, \overline{\beta})$ satisfy
the monotonicity condition (\ref{monqr}),
\end{itemize}

then 
\begin{equation*}
\alpha=\overline{\alpha}, \; \beta=\overline{\beta}, \; U=\overline{U}.
\end{equation*}
\end{prop}

\begin{proof}
Let us define for every $t\in [0,1]$ 
\begin{equation*}
\varphi(t):=\int_0^t \alpha(s)ds, \; b(t):=\int_0^t \beta(s)ds.
\end{equation*}
Let us also define for $(x,y)$ in ${\mathbb{R}}^{N+1}$: 
\begin{equation*}
\psi(x,y):=\max_{t\in [0,1]} \{ty-\varphi(t)-b(t)^{\top} x\}
\end{equation*}
thanks to the monotonicity condition (\ref{monqr}), the maximization program
above is strictly concave in $t$ for every $y$ and $m$-a.e.$x$. We then
remark that $Y=\alpha(U)+\beta(U)^{\top} X=\varphi^{\prime
}(U)+b^{\prime}(U)^{\top}X$ exactly is the first-order condition for the
above maximization problem when $(x,y)=(X,Y)$. In other words, we have 
\begin{equation}  \label{ineqq}
\psi(x,y)+b(t)^{\top}x + \varphi(t)\ge ty, \; \forall (t,x,y)\in [0,1]\times 
{\mathbb{R}}^N\times {\mathbb{R}}
\end{equation}
with an equality for $(x,y,t)=(X,Y,U)$ i.e. 
\begin{equation}  \label{eqas}
\psi(X,Y)+b(U)^{\top} X + \varphi(U)=UY, \; \mbox{ almost surely. }
\end{equation}
Using the fact that ${\mathscr{Law}}(U)={\mathscr{Law}}(\overline{U})$ and
the fact that mean-independence gives ${\mathbb{E}}(b(U)^{\top} X)={\mathbb{E%
}}(b(\overline{U})^{\top} X)=0$, we have 
\begin{equation*}
{\mathbb{E}}(UY)={\mathbb{E}}( \psi(X,Y)+b(U)^{\top} X + \varphi(U))= {%
\mathbb{E}}( \psi(X,Y)+b(\overline{U})^{\top} X + \varphi(\overline{U})) \ge 
{\mathbb{E}}(\overline{U} Y)
\end{equation*}
but reversing the role of $U$ and $\overline{U}$, we also have ${\mathbb{E}}%
(UY)\le {\mathbb{E}}(\overline{U} Y)$ and then 
\begin{equation*}
{\mathbb{E}}(\overline{U} Y)= {\mathbb{E}}( \psi(X,Y)+b(\overline{U})^{\top}
X + \varphi(\overline{U}))
\end{equation*}
so that, thanks to inequality (\ref{ineqq}) 
\begin{equation*}
\psi(X,Y)+b(\overline{U})^{\top} X + \varphi(\overline{U})=\overline{U} Y,
\; \mbox{ almost surely }
\end{equation*}
which means that $\overline{U}$ solves $\max_{t\in [0,1]}
\{tY-\varphi(t)-b(t)^{\top} X\}$ which, by strict concavity admits $U$ as
unique solution. This proves that $U=\overline{U}$ and thus 
\begin{equation*}
\alpha(U)-\overline{\alpha}(U)=(\overline{\beta}(U)-\beta(U))^{\top} X
\end{equation*}
taking the conditional expectation with respect to $U$ on both sides we then
obtain $\alpha=\overline{\alpha}$ and thus $\beta(U)^{\top} X=\overline{\beta%
}(U)^{\top} X$ almost surely. We then compute 
\begin{equation*}
\begin{split}
F(x, \alpha(t)+\beta(t)^{\top} x)&= \mathbb{P}(\alpha(U)+\beta(U)^{\top} X
\le \alpha(t)+\beta(t)^{\top} x \vert X=x) \\
&=\mathbb{P}( \alpha(U)+ \beta(U)^{\top} x \le \alpha(t)+\beta(t)^{\top} x
\vert X=x) \\
&=\mathbb{P}(U\le t \vert X=x)
\end{split}%
\end{equation*}
and similarly $F(x, \alpha(t)+\overline{\beta}(t)^{\top} x)=\mathbb{P}(U\le
t \vert X=x)=F(x, \alpha(t)+\beta(t)^{\top} x)$. Thanks to (\ref{hypo1}), we
deduce that $\beta(t)^{\top} x=\overline{\beta}(t)^{\top} x$ for $m$-a.e. $x$
and every $t\in[0,1]$. Finally, the previous considerations and the
nondegeneracy condition (\ref{nondegX}) enable us to conclude that $\beta=%
\overline{\beta}$.
\end{proof}

\begin{coro}
Assume (\ref{hypo1})-(\ref{finitemoment})-(\ref{normalizX})-(\ref{nonvert})
and (\ref{nondegX}). If quantile regression is quasi-specified, the
regression coefficients $(\alpha^{QR}, \beta^{QR})$ are uniquely defined and
if $Y$ can be written as 
\begin{equation*}
Y=\alpha(U)+\beta(U)^{\top} X
\end{equation*}
for $U$ uniformly distributed, $X$ being mean independent from $U$, $%
(\alpha, \beta)$ continuous such that the monotonicity condition (\ref{monqr}%
) holds then necessarily 
\begin{equation*}
\alpha=\alpha^{QR}, \; \beta=\beta^{QR}.
\end{equation*}
\end{coro}

To sum up, we have shown that quasi-specification is equivalent to the
validity of the factor linear model: 
\begin{equation*}
Y=\alpha(U)+\beta(U)^{\top} X
\end{equation*}
for $(\alpha, \beta)$ continuous and satisfying the monotonicity condition (%
\ref{monqr}) and $U$, uniformly distributed and such that $X$ is
mean-independent from $U$. This has to be compared with the decomposition of
paragraph \ref{subsec-cond-quant} where $U$ is required to be independent
from $X$ but the dependence of $Y$ with respect to $U$, given $X$, is given
by a nondecreasing function of $U$ which is not necessarily affine in $X$.

\section{Quantile regression without specification}

Now we wish to address quantile regression in the case where neither
specification nor quasi-specification can be taken for granted. In such a
general situation, keeping in mind the remarks from the previous paragraphs,
we can think of two natural approaches.

The first one consists in studying directly the correlation maximization
with a mean-independence constraint (\ref{maxcorrmi}). The second one
consists in getting back to the Koenker and Bassett $t$ by $t$ problem (\ref%
{dt}) but adding as an additional global consistency constraint that $V_t$
should be nonincreasing (which we abbreviate as $V_t \downarrow$) with
respect to $t$:

\begin{equation}  \label{monconstr}
\sup\{{\mathbb{E}}(\int_0^1 V_t Ydt ) \; : \: V_t \downarrow, \; V_t\in
[0,1],\; {\mathbb{E}}(V_t)=(1-t), \; {\mathbb{E}}(V_t X)=0\}
\end{equation}

Our aim is to compare these two approaches (and in particular to show that
the maximization problems (\ref{maxcorrmi}) and (\ref{monconstr}) have the
same value) as well as their dual formulations. Before going further, let us
remark that (\ref{maxcorrmi}) can directly be considered in the multivariate
case whereas the monotonicity constrained problem (\ref{monconstr}) makes
sense only in the univariate case.

As proven in Carlier et al. (2016), (\ref{maxcorrmi}) is dual to 
\begin{equation}  \label{dualmi}
\inf_{(\psi, \varphi, b)} \{{\mathbb{E}}(\psi(X,Y))+{\mathbb{E}}(\varphi(U))
\; : \; \psi(x,y)+ \varphi(u)\ge uy -b(u)^{\top} x\}
\end{equation}
which can be reformulated as: 
\begin{equation}  \label{dualmiref}
\inf_{(\varphi, b)} \int \max_{t\in [0,1]} ( ty- \varphi(t) -b(t)^{\top} x)
\nu(dx, dy) +\int_0^1 \varphi(t) dt
\end{equation}
in the sense that\footnote{%
With a little abuse of notations when a reference number (A) refers to a
maximization (minimization) problem, we will simply write $\sup(A)$ ($\inf(A)
$) to the denote the value of this optimization problem.} 
\begin{equation}  \label{nodualgap}
\sup (\ref{maxcorrmi})=\inf(\ref{dualmi})=\inf (\ref{dualmiref}).
\end{equation}

The existence of a solution to (\ref{dualmi}) is not straightforward and is
established under appropriate assumptions in Carlier et al. (2017) in the
multivariate case. The following result shows that there is a $t$-dependent
reformulation of (\ref{maxcorrmi}):

\begin{lem}
\label{treform} The value of (\ref{maxcorrmi}) coincides with 
\begin{equation}  \label{monconstr01}
\sup\{{\mathbb{E}}(\int_0^t V_t Ydt ) \; : \: V_t \downarrow, \; V_t\in
\{0,1\},\; {\mathbb{E}}(V_t)=(1-t), \; {\mathbb{E}}(V_t X)=0\}.
\end{equation}
\end{lem}

\begin{proof}
Let $U$ be admissible for (\ref{maxcorrmi}) and define $V_t:=\mathbf{1}%
_{\{U>t\}}$ then $U=\int_0^1 V_t dt$ and obviously $(V_t)_t$ is admissible
for (\ref{monconstr01}), we thus have $\sup (\ref{maxcorrmi}) \le \sup (\ref%
{monconstr01})$. Take now $(V_t)_t$ admissible for (\ref{monconstr01}) and
let $V:=\int_0^1 V_t dt$, we then have 
\begin{equation*}
V>t \Rightarrow V_t=1\Rightarrow V\ge t
\end{equation*}
since ${\mathbb{E}}(V_t)=(1-t)$ this implies that $V$ is uniformly
distributed and $V_t=\mathbf{1}_{\{V>t\}}$ almost surely so that ${\mathbb{E}%
}(X \mathbf{1}_{\{V>t\}})=0$ which implies that $X$ is mean-independent from 
$V$ and thus ${\mathbb{E}}(\int_0^1 V_t Y dt)\le \sup (\ref{maxcorrmi})$. We
conclude that $\sup (\ref{maxcorrmi}) = \sup (\ref{monconstr01})$.
\end{proof}

Let us now define 
\begin{equation*}
{\mathcal{C}}:=\{v \; : \; [0,1]\mapsto [0,1], \; \downarrow \}
\end{equation*}

Let $(V_t)_t$ be admissible for (\ref{monconstr}) and set 
\begin{equation*}
v_t(x,y):={\mathbb{E}}(V_t \vert X=x, Y=y), \; V_t:= v_t(X,Y)
\end{equation*}
it is obvious that $(V_t)_t$ is admissible for (\ref{monconstr}) and by
construction ${\mathbb{E}}(V_t Y)={\mathbb{E}}(V_t Y)$. Moreover the
deterministic function $(t,x,y)\mapsto v_t(x,y)$ satisfies the following
conditions: 
\begin{equation}  \label{CCt}
\mbox{for fixed $(x,y)$, } t\mapsto v_t(x,y) \mbox{ belongs to $\CC$,}
\end{equation}
and for a.e. $t\in [0,1]$, 
\begin{equation}  \label{moments}
\int v_t(x,y) \nu(dx, dy)=(1-t), \; \int v_t(x,y) x\nu(dx, dy)=0.
\end{equation}
Conversely, if $(t,x,y)\mapsto v_t(x,y)$ satisfies (\ref{CCt})-(\ref{moments}%
), $V_t:=v_t(X,Y)$ is admissible for (\ref{monconstr}) and ${\mathbb{E}}(V_t
Y)=\int v_t(x,y) y \nu(dx, dy)$. All this proves that $\sup(\ref{monconstr})$
coincides with 
\begin{equation}  \label{supvt}
\sup_{(t,x,y)\mapsto v_t(x,y)} \int v_t(x,y) y \nu(dx, dy)dt 
\mbox{ subject
to: } (\ref{CCt})-(\ref{moments})
\end{equation}

\begin{thm}
\label{equivkb} The shape constrained quantile regression problem (\ref%
{monconstr}) is related to the correlation maximization with a mean
independence constraint (\ref{maxcorrmi}) by: 
\begin{equation*}
\sup (\ref{maxcorrmi})=\sup (\ref{monconstr}).
\end{equation*}
\end{thm}

\begin{proof}
We know from lemma \ref{treform} and the remarks above that 
\begin{equation*}
\sup (\ref{maxcorrmi})=\sup (\ref{monconstr01}) \le \sup (\ref{monconstr}%
)=\sup (\ref{supvt}).
\end{equation*}
We now get rid of constraints (\ref{moments}) by rewriting (\ref{supvt}) in
sup-inf form as 
\begin{equation*}
\begin{split}
\sup_{\quad\text{$v_t$ satisfies (\ref{CCt})}\quad} \inf_{(\alpha, \beta)}
\int v_t(x,y)(y-\alpha(t)-\beta(t)^{\top} x) \nu(dx,dy)dt +\int_0^1
(1-t)\alpha(t) dt
\end{split}%
\end{equation*}
Recall that one always have $\sup \inf \le \inf \sup$ so that $\sup(\ref%
{supvt})$ is less than 
\begin{equation*}
\begin{split}
\inf_{(\alpha, \beta)} \sup_{\quad\text{$v_t$ satisf. (\ref{CCt})}\quad}
\int v_t(x,y)(y-\alpha(t)-\beta(t)^{\top} x) \nu(dx,dy)dt +\int_0^1
(1-t)\alpha(t) dt \\
\le \inf_{(\alpha, \beta)} \int \Big (\sup_{v\in {\mathcal{C}}} \int_0^1
v(t)(y-\alpha(t)-\beta(t)^{\top}x)dt \Big) \nu(dx,dy)+ \int_0^1
(1-t)\alpha(t) dt
\end{split}%
\end{equation*}
It follows from Lemma \ref{suppC} below that, for $q\in L^1(0,1)$ defining $%
Q(t):=\int_0^t q(s) ds$, one has 
\begin{equation*}
\sup_{v\in {\mathcal{C}}} \int_0^1 v(t) q(t)dt=\max_{t\in [0,1]} Q(t).
\end{equation*}
So setting $\varphi(t):=\int_0^t \alpha(s) ds$, $b(t):=\int_0^t \beta(s)ds$
and remarking that integrating by parts immediately gives 
\begin{equation*}
\int_0^1 (1-t)\alpha(t) dt=\int_0^1 \varphi(t) dt,
\end{equation*}
we have 
\begin{equation*}
\begin{split}
\sup_{v\in {\mathcal{C}}} \int_0^1 v(t)(y-\alpha(t)-\beta(t)^{\top}x)dt +
\int_0^1 (1-t)\alpha(t) dt \\
= \max_{t\in[0,1]} \{t y-\varphi(t)-b(t)^{\top} x\} +\int_0^1 \varphi(t) dt.
\end{split}%
\end{equation*}
This yields%
\begin{equation*}
\sup(\ref{supvt}) \le \inf_{(\varphi, b)} \int \max_{t\in [0,1]} ( ty-
\varphi(t) -b(t)^{\top} x) \nu(dx, dy) +\int_0^1 \varphi(t) dt =\inf (\ref%
{dualmiref})
\end{equation*}
but we know from (\ref{nodualgap}) that $\inf (\ref{dualmiref}) =\sup (\ref%
{maxcorrmi})$ which ends the proof.
\end{proof}

In the previous proof, we have used the elementary result (proven in the
appendix)

\begin{lem}
\label{suppC} Let $q\in L^1(0,1)$ and define $Q(t):=\int_0^t q(s) ds$ for
every $t\in [0,1]$, one has 
\begin{equation*}
\sup_{v\in {\mathcal{C}}} \int_0^1 v(t) q(t)dt=\max_{t\in [0,1]} Q(t).
\end{equation*}
\end{lem}

\section{Vector quantiles, vector quantile regression and optimal transport}

We now consider the case where $Y$ is a random vector with values in ${%
\mathbb{R}}^d$ with $d\geq 2$. The notion of quantile does not have an
obvious generalization in the multivariate setting however, the various
correlation maximization problems we have encountered in the previous
sections still make sense (provided $Y$ is integrable say) in dimension $d$
and are related to optimal transport theory. The aim of this section is to
briefly summarize the optimal transport approach to quantile regression
introduced in Carlier et al. (2016) and further analyzed in their follow-up
2017 paper.



\subsection{Brenier's map as a vector quantile}

From now on we fix as a reference measure the uniform measure on the unit
cube $[0,1]^d$ i.e. 
\begin{equation}
\mu_d:={\mathcal{U}}([0,1]^d)
\end{equation}
Given $Y$, an integrable ${\mathbb{R}}^d$-valued random variable on $%
(\Omega, {\mathcal{F}}, \mathbb{P})$, a remarkable theorem due to Brenier
(1991) and extended by McCann (1995) implies that there exists a unique $%
U\sim \mu_d$ and a unique (up to the addition of a constant) convex function
defined on $[0,1]^d$ such that 
\begin{equation}  \label{factpol}
Y=\nabla \varphi(U).
\end{equation}
The map $\nabla \varphi$ is called the Brenier's map between $\mu_d$ and ${%
\mathscr{Law}}(Y)$.

The convex function $\varphi$ is not necessarily differentiable but being
convex it is differentiable at Lebesgue-a.e. point of $[0,1]^d$ so that $%
\nabla \varphi(U)$ is well defined almost surely, it is worth at this point
recalling that the Legendre transform of $\varphi$ is the convex function: 
\begin{equation}
\varphi^*(y):=\sup_{u\in [0,1]^d} \{ u^{\top} y -\varphi(u)\}
\end{equation}
and that the subdifferentials of $\varphi$ and $\varphi^*$ are defined
respectively by 
\begin{equation*}
\partial \varphi(u):=\{y \in {\mathbb{R}}^d \; : \;
\varphi(u)+\varphi^*(y)=u^{\top} y\}
\end{equation*}
and 
\begin{equation*}
\partial \varphi^*(y):=\{u \in [0,1]^d \; : \;
\varphi(u)+\varphi^*(y)=u^{\top} y\}
\end{equation*}
so that $\partial \varphi$ and $\partial \varphi^*$ are inverse to each
other in the sense that 
\begin{equation*}
y\in \partial \varphi(u) \Leftrightarrow u\in \partial \varphi^*(y)
\end{equation*}
which is often refered to in convex analysis as the Fenchel reciprocity
formula\footnote{%
Note the analogy with the fact that in the univariate case the cdf and the
quantile of $Y$ are generalized inverse to each other.}. Note then that (\ref%
{factpol}) implies that 
\begin{equation*}
U\in \partial \varphi^*(Y) \mbox{ almost surely}.
\end{equation*}
If both $\varphi$ and $\varphi^*$ are differentiable, their subgradients
reduce to the singleton formed by their gradient and the Fenchel
recirprocity formula simply gives $\nabla \varphi^{-1}=\nabla \varphi^*$.
Recalling the subgradient of the convex function $\varphi$ is monotone in
the sense that whenever $y_1\in \partial \varphi(u_1)$ and $y_2\in \partial
\varphi(u_2)$ one has 
\begin{equation*}
(y_1-y_2)^{\top} (u_1-u_2)\geq 0,
\end{equation*}
we see that gradients of convex functions are a genelarization to the
multivariate case of monotone univariate maps. It is therefore natural in
view of (\ref{factpol}) to define the vector quantile of $Y$ as:

\begin{defi}
The vector quantile of $Y$ is the Brenier's map between $\mu_d$ and ${%
\mathscr{Law}}(Y)$.
\end{defi}

Now, it is worth noting that the Brenier's map (and the uniformly
distributed random vector $U$ in (\ref{factpol})) are not abstract objects,
they have a variational characterization related to optimal transport%
\footnote{%
In the case where ${\mathbb{E}}(\Vert Y\Vert^2)<+\infty$, (\ref{mmmc}) is
equivalent to minimize ${\mathbb{E}}(\Vert V- Y\Vert^2)$ among uniformly
distributed $V$'s.}. Consider indeed 
\begin{equation}  \label{mmmc}
\sup\{ {\mathbb{E}}(V^{\top} Y) \; : \; V\sim \mu_d\}
\end{equation}
and its dual 
\begin{equation}  \label{mmmcdual}
\inf_{f, g} \{\int_{[0,1]^d} f \mbox{d} \mu_d + {\mathbb{E}}(g(Y)) \; : \;
f(u)+g(y) \geq u^{\top} y, \; \forall (u,y)\in [0,1]^d\times {\mathbb{R}}^d\}
\end{equation}
then $U$ in (\ref{factpol}) is the unique solution of (\ref{mmmc}) and any
solution $(f,g)$ of the dual (\ref{mmmcdual}) satisfies $\nabla f=\nabla
\varphi$ $\mu_d$-a.e..

\subsection{Conditional vector quantiles}

Assume now as in paragraph \ref{subsec-cond-quant} that we are also given a
random vector $X\in {\mathbb{R}}^N$. As in (\ref{notationslawxy}), we denote
by $\nu$ the law of $(X,Y)$, by $m$ the law of $X$ and by $\nu(.\vert x)$
the conditional law of $Y$ given $X=x$ (the only difference with (\ref%
{notationslawxy}) is that $Y$ is ${\mathbb{R}}^d$-valued). Conditional
vector quantile are then defined as

\begin{defi}
For $m={\mathscr{Law}}(X)$-a.e. $x\in {\mathbb{R}}^N$, the vector
conditional quantile of $Y$ given $X=x$ is the Brenier's map between $\mu_d:=%
{\mathcal{U}}([0,1]^d)$ and $\nu(.\vert x):={\mathscr{Law}}(Y\vert X=x)$. We
denote this well defined map as $\nabla \varphi_x$ where $\varphi_x$ is a
convex function on $[0,1]^d$.
\end{defi}

If both $\varphi_x$ and its Legendre transform 
\begin{equation*}
\varphi_x^*(y):=\sup_{u\in [0,1]^d} \{u^{\top} y-\varphi_x(u)\}
\end{equation*}
are differentiable\footnote{%
A deep regularity theory initated by Caffarelli (1992) in the 1990's gives
conditions on $\nu(.\vert x)$ such that this is in fact the case that the
optimal transport map is smooth and/or invertible, we refer the interested
reader to the textbook of Figalli (2017) for a detailed and recent account
of this regularity theory.}, one can define the random vector: 
\begin{equation*}
U:=\nabla \varphi_X^*(Y)
\end{equation*}
which is equivalent to 
\begin{equation}  \label{polarmdi}
Y=\nabla \varphi_X(U).
\end{equation}
One can check exactly as in the proof of Proposition \ref{univaropt} for the
univariate case that if $Y$ is integrable then 
\begin{equation*}
U\sim \mu_d, \; U\perp \! \! \! \perp X
\end{equation*}
and $U$ solves 
\begin{equation}  \label{maxcorrindep}
\max \{{\mathbb{E}}(V^{\top} Y), \; V\sim \mu_d, \; V\perp \! \! \! \perp
X\}.
\end{equation}

\subsection{Vector quantile regression}

When one assumes that the convex function $\varphi_x$ is affine with respect
to the explanatory variables $x$ (specification): 
\begin{equation*}
\varphi_x(u)=\varphi(u)+ b(u)^{\top} x
\end{equation*}
with $\varphi$ : $[0,1]^d \to {\mathbb{R}}$ and $b$ : $[0,1]^d \to {\mathbb{R%
}}^N$ smooth, the conditional quantile is itself affine and the relation (%
\ref{polarmdi}) takes the form 
\begin{equation}  \label{funcformreg}
Y=\nabla \varphi_X(U)=\alpha(U)+ \beta(U)X, \mbox{ for } \alpha =\nabla
\varphi, \; \beta:=Db^{\top}.
\end{equation}
This affine form moreover implies that not only $U$ maximizes the
correlation with $Y$ among uniformly distributed random vectors independent
from $X$ but in the larger class of uniformly distributed random vectors for
which\footnote{%
here we assume that both $X$ and $Y$ are integrable} 
\begin{equation*}
{\mathbb{E}}(X\vert U)={\mathbb{E}}(X)=0.
\end{equation*}
This is the reason why the study of 
\begin{equation}  \label{maxcorrmeanindep}
\max \{{\mathbb{E}}(V^{\top} Y), \; V\sim \mu_d, \; {\mathbb{E}}(X\vert
V)=0\}
\end{equation}
is the main tool in the approach of Carlier et al. (2016, 2017) to vector
quantile regression. Let us now briefly summarize the main findings in these
two papers. First observe that (\ref{maxcorrmeanindep}) can be recast as a
linear program by setting $\pi:={\mathscr{Law}}(U, X,Y)$ and observing that $%
U$ solves (\ref{maxcorrmeanindep}) if and only if $\pi$ solves 
\begin{equation}  \label{VQRprimal}
\max_{\pi\in \mathop{\mathrm{MI}}\nolimits( \mu_d, \nu)} \int_{
[0,1]^d\times {\mathbb{R}}^N\times {\mathbb{R}}^d} u^{\top} y \mbox{d}
\pi(u,x,y)
\end{equation}
where $\mathop{\mathrm{MI}}\nolimits(\nu, \mu)$ is the set of probability
measures which satisfy the linear constraints:

\begin{itemize}
\item the first marginal of $\pi$ is $\mu_d$, i.e., for every $\varphi \in
C([0,1]^d, {\mathbb{R}})$: 
\begin{equation*}
\int_{ [0,1]^d \times {\mathbb{R}}^N\times {\mathbb{R}}^d} \varphi(u)\mbox{d}
\pi(u,x,y)=\int_{[0,1]^d} \varphi(u) \mbox{d} \mu_d(u),
\end{equation*}

\item the second marginal of $\pi$ is $\nu$, i.e., for every $\psi \in C_b({%
\mathbb{R}}^N\times {\mathbb{R}}^d, {\mathbb{R}})$: 
\begin{equation*}
\begin{split}
\int_{[0,1]^d \times{\mathbb{R}}^N\times {\mathbb{R}}^d} \psi(x,y)\mbox{d}
\pi(u,x,y)&=\int_{{\mathbb{R}}^N\times {\mathbb{R}}^d} \psi(x,y) \mbox{d}
\nu(x,y) \\
& ={\mathbb{E}}(\psi(X,Y)),
\end{split}%
\end{equation*}

\item the conditional expectation of $x$ given $u$ is $0$, i.e., for every $%
b\in C([0,1]^d, {\mathbb{R}}^N)$: 
\begin{equation*}
\int_{[0,1]^d \times{\mathbb{R}}^N\times {\mathbb{R}}^d} b(u)^{\top} x %
\mbox{d} \pi(u,x,y)=0.
\end{equation*}
\end{itemize}

The dual of the linear program (\ref{maxcorrmeanindep}) then reads 
\begin{equation}  \label{dualmci}
\inf_{(\varphi,\psi,b)} \int_{[0,1]^d} \varphi \mbox{d} \mu_d+ \int_{{%
\mathbb{R}}^N\times {\mathbb{R}}^d} \psi(x,y) \mbox{d} \nu(x,y)
\end{equation}
subject to the pointwise constraint 
\begin{equation*}
\varphi(u)+b(u)^{\top} x+\psi(x,y) \geq u^{\top} y
\end{equation*}
given $b$ and $\varphi$ the lowest $\psi$ fitting this constraint being the
(convex in $y$) function 
\begin{equation*}
\psi(x,y):=\sup_{u\in [0,1]^d} \{ u^{\top} y-\varphi(u)-b(u)^{\top} x\}.
\end{equation*}
The existence of a solution $(\psi, \varphi, b)$ to (\ref{dualmci}) is
established in Carlier et al. (2016) (under some assumptions on $\nu$) and
optimality for $U$ in (\ref{maxcorrmeanindep}) is characterized by the
pointwise complementary slackness condition 
\begin{equation*}
\varphi(U)+b(U)^{\top} X+\psi(X,Y) = U^{\top} Y \mbox{ almost surely}.
\end{equation*}
If $\varphi$ and $b$ were smooth we could deduce from the latter that 
\begin{equation*}
Y=\nabla \varphi(U)+Db(U)^{\top} U=\nabla \varphi_X(U), \; \mbox{ for }
\varphi_x(u):=\varphi(u)+b(u)^{\top} x
\end{equation*}
which is exactly (\ref{funcformreg}). So specification of vector quantile
regression is essentially the same as assuming this smoothness and the
convexity of $u\mapsto \varphi_x(u):=\varphi(u)+b(u)^{\top} x$. In general,
these properties cannot be taken for granted and what can be deduced from
complementary slackness is given by the weaker relations 
\begin{equation*}
\varphi_X(U)=\varphi_X^{**}(U), \; Y\in \partial \varphi_X^{**}(U) 
\mbox{
almost surely,}
\end{equation*}
were $\varphi_x^{**}$ is the convex envelope of $\varphi_x$ (i.e. the
largest convex function below $\varphi_x$), we refer the reader to Carlier
et al. (2017) for details.

\section{Discretization, regularization, numerical minimization}

\subsection{Discrete optimal transport with a mean independence constraint}

We now turn to a discrete setting for implementation purposes, and consider
data $(X_j,Y_j)_{j=1..J}$ distributed according to the empirical measure $%
\nu=\sum_{j=1}^J \nu_j \delta_{(x_j,y_j)}$, and a $[0,1]^d$-uniform sample $%
(U_i)_{i=1, \ldots, I}$ with empirical measure $\mu=\sum_{i=1}^I \mu_i
\delta_{u_i}$. In this setting, the vector quantile regression primal (\ref%
{VQRprimal}) writes 
\begin{equation*}
\max_{\pi \in \mathbb{R}_{+}^{I\times J}} \sum_{i=1}^I \sum_{j=1}^J
u_i^{\top} y_j \pi_{ij}
\end{equation*}
subject to marginal constraints $\forall j, \sum_{i} \pi_{ij} = \nu_j$ and $%
\forall i, \sum_j \pi_{ij} = \mu_i$ and the mean-independence constraint
between $X$ and $U$: $\forall i, \sum_j x_j \pi_{ij}=0$. Its dual
formulation (\ref{dualmci}) reads 
\begin{equation*}
\inf_{(\phi_i)_i,(\psi_j)_j,(b_i)_i} \sum_{j=1}^J \psi_j \nu_j +
\sum_{i=1}^I \phi_i \mu_i
\end{equation*}
subject to the constraint 
\begin{equation*}
\forall i, j, \phi_i + b_i^{\top} x_j + \psi_j \geq u_i^{\top} y_j.
\end{equation*}

\subsection{The Regularized Vector Quantile Regression (RVQR)\ problem\label%
{par:regul}}

Using the optimality condition $\phi
_{i}=\max_{j}u_{i}^{\top}y_{j}-b_{i}^{\top}x_{j}-\psi _{j}$, we obtain the
unconstrained formulation 
\begin{equation*}
\inf_{(\psi _{j})_{j},(b_{i})_{i}}\sum_{j}\psi _{j}\nu _{j}+\sum_{i}\mu
_{i}\left( \max_{j}u_{i}^{\top}y_{j}-b_{i}^{\top}x_{j}-\psi _{j}\right) .
\end{equation*}%
Replacing the maximum with its smoothed version\footnote{%
Recall that the softmax with regularization parameter $\eps>0$ of $(\alpha
_{1},\ldots ,\alpha _{J})$ is given by ${\mathrm{Softmax}}_{\eps}(\alpha
_{1},\ldots \alpha _{J}):=\eps\log (\sum_{j=1}^{J}e^{\frac{\alpha _{j}}{\eps}%
})$.}, given a small regularization parameter $\varepsilon $, yields the
smooth convex minimization problem (see Cuturi and Peyré (2016) for more
details in connection with entropic regularization of optimal transport),
which we call the \emph{Regularized Vector Quantile Regression} (RVQR)\
problem%
\begin{equation}
\inf_{(\psi _{j})_{j},(b_{i})_{i}}J(\psi ,b):=\sum_{j}\psi _{j}\nu
_{j}+\varepsilon \sum_{i}\mu _{i}\log \left( \sum_{j}\exp \left( \frac{1}{%
\varepsilon }[u_{i}^{\top}y_{j}-b_{i}^{\top}x_{j}-\psi _{j}]\right) \right) .
\label{smootheddual}
\end{equation}

We then have the following duality result\footnote{%
Which can be proved either by using the Fenchel-Rockafellar duality theorem
or by hand. Indeed, in the primal, there are only finitely many linear
constraints and nonnegativity constraints are not binding because of the
entropy. The existence of Lagrange multipliers for the equality constraints
is then straightforward.}:

\begin{thm}
\label{thm:dualityRegularized}The RVQR problem%
\begin{eqnarray*}
\max_{\pi _{ij}\geq 0} &&\sum_{ij}\pi _{ij}\left( u_{i}^{\top}y_{j}\right)
-\varepsilon \sum_{ij}\pi _{ij}\log \pi _{ij} \\
&&\sum_{j}\pi _{ij}=\mu _{i} \\
&&\sum_{i}\pi _{ij}=\nu _{j} \\
&&\sum_{j}\pi _{ij}x_{j}=\sum_{j}\nu _{j}x_{j}
\end{eqnarray*}%
has dual~(\ref{smootheddual}), or equivalently%
\begin{equation*}
\min_{\varphi _{i},v_{j}}\sum_{i}\mu _{i}\varphi _{i}+\sum_{j}\psi _{j}\nu
_{j}+\varepsilon \sum_{ij}\exp \left( \frac{1}{\varepsilon }%
[u_{i}^{\top}y_{j}-\varphi _{i}-b_{i}^{\top}x_{j}-\psi _{j}]\right) .
\end{equation*}
\end{thm}

Note that the objective $J$ in~(\ref{smootheddual}) remains invariant under
the two transformations

\begin{itemize}
\item $(b,\psi) \leftarrow (b+c, \psi -c^{\top} x)$ with $c\in {\mathbb{R}}^N
$ is a constant translation vector,

\item $\psi \leftarrow \psi+\lambda$ where $\lambda\in {\mathbb{R}}$ is a
constant.
\end{itemize}

These two invariances enable us to fix the value of $b_1=0$ and (for
instance) to chose $\lambda$ in such a way that $\sum_{i,j} \exp \left( 
\frac{1}{\varepsilon} [u_i^{\top} y_j - b_i^{\top} x_j - \psi_j] \right))=1$.

\noindent \emph{Remark.} This formulation is eligible for stochastic
optimization techniques when the number of $(X,Y)$ observations is very
large. Stochastic optimization w.r.t. $\psi$ can be performed using the
stochastic averaged gradient algorithm, see Genevay et al. (2016),
considering the equivalent objective 
\begin{equation*}
\inf_{\psi, \phi, b} \sum_j h_\varepsilon(x_j,y_j,\psi,\phi,b)\nu_j
\end{equation*}
with $h_\varepsilon(x_j,y_j,\psi,\phi,b)=\psi_j+\sum_i \mu_i \phi_i +
\varepsilon\sum_i \exp \left( \frac{1}{\varepsilon} [u_i^{\top} y_j -
b_i^{\top} x_j - \psi_j - \phi_i] \right)$. Such techniques are not needed
to compute $b$ since the number of $U$ samples (i.e. the size of $b$) is set
by the user.

\subsection{Gradient descent}

As already noted the objective $J$ in (\ref{smootheddual}) is convex%
\footnote{%
it is even strictly convex once we have chosen normalizations which take
into account the two invariances of $J$ explained above.} and smooth. Its
gradient has the explicit form 
\begin{equation}  \label{derivinpsi}
\frac{\partial J}{\partial \psi_j}:=\nu_j-\sum_{i=1}^I \mu_i \frac{
e^{\theta_{ij}} } {\sum_{k=1}^J e^{\theta_{ik} } } \mbox{ where }
\theta_{ij}=\frac{1}{\varepsilon} [u_i^{\top} y_j - b_i^{\top} x_j - \psi_j]
\end{equation}
and 
\begin{equation}  \label{derivinb}
\frac{\partial J}{\partial b_i}:=- \mu_i \frac{\sum_{k=1}^J x_k
e^{\theta_{ik}} } {\sum_{k=1}^J e^{\theta_{ik}}}.
\end{equation}
To solve (\ref{smootheddual}) numerically, we therefore can use a gradient
descent mehod. An efficient way to do it is to use Nesterov accelerated
gradient algorithm see Nesterov (1983) and Beck and Teboulle (2009). Note
that if $\psi, b $ solves (\ref{smootheddual}), the fact that the partial
derivatives in (\ref{derivinpsi})-(\ref{derivinb}) vanish imply that the
coupling 
\begin{equation*}
\alpha^\eps_{ij}:=\mu_i \frac{ e^{\theta_{ij}} } {\sum_{k=1}^J
e^{\theta_{ik} } }
\end{equation*}
satisfies the constraint of fixed marginals and mean-independence of the
primal problem. Since the index $j$ corresponds to observations it is
convenient to introduce for every $x\in {\mathcal{X}}:=\{x_1, \ldots, x_J\}$
and $y\in {\mathcal{Y}}:=\{y_1, \ldots y_j\}$ the probability 
\begin{equation*}
\pi^\eps(x,y, u_i):=\sum_{j \; : \; x_j=x, \; y_j=y} \alpha^\eps_{ij}.
\end{equation*}


\section{Results}

{\textbf{Quantiles computation.}} The discrete probability $\pi^\eps$ is an
approximation (because of the regularization $\eps$) of ${\mathscr{Law}}%
(U,X,Y)$ where $U$ solves (\ref{maxcorrmeanindep}). The corresponding
approximate quantile $Q^\eps_X(U)$ is given by ${\mathbb{E}}_{\pi^\eps}
[Y|X,U]$. In the above discrete setting, this yields 
\begin{equation*}
Q^\eps_x(u_i):={\mathbb{E}}_{\pi^\eps} [Y|X=x,U=u_i] =\sum_{y\in {\mathcal{Y}%
}} y \frac{\pi^\eps(x,y,u_i) }{ \sum_{y^{\prime }\in {\mathcal{Y}}} \pi^\eps%
(x,y^{\prime },u_i) }.
\end{equation*}

\noindent \emph{Remark.} To estimate the conditional distribution of $Y$
given $U=u$ and $X=x$, we can use kernel methods. In the experiments, we
compute approximate quantiles as means on neighborhoods of $X$ values to
make up for the lack of replicates. This amounts to considering ${\mathbb{E}}%
_{\pi ^{\eps}}[Y|X\in B_{\eta }(x),U=u_{i}]$ where $B_{\eta }(x)$ is a
Euclidean ball of radius $\eta $ centered on $x$.

{\textbf{Empirical illustrations.}} We demonstrate the use of this approach
on a series of health related experiments. We use the ``ANSUR II'' dataset
(Anthropometric Survey of US Army Personnel), which can be found online%
\footnote{%
https://www.openlab.psu.edu/ansur2/}. This dataset is one of the most
comprehensive publicly available data sets on body size and shape,
containing 93 measurements for over 4,082 male adult US military personnel.
It allows us to easily build multivariate dependent variables.

\begin{figure}[]
\centering
\begin{minipage}{.5\textwidth}
  \centering
  \includegraphics[scale=0.4]{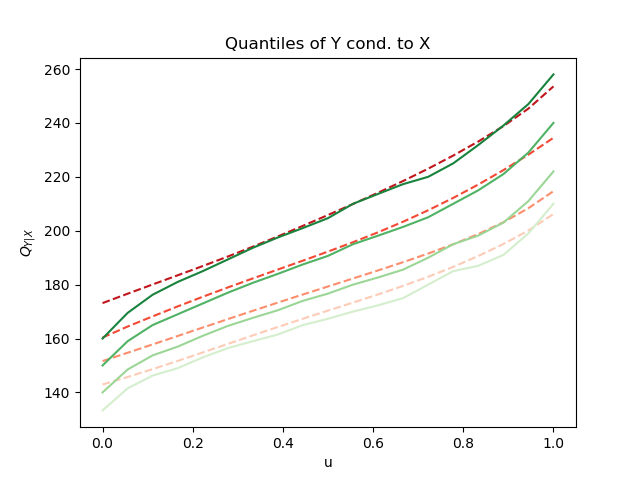}
  \caption*{First dimension}
\end{minipage}\hspace*{0.01cm} 
\begin{minipage}{.5\textwidth}
  \centering
  \includegraphics[scale=0.4]{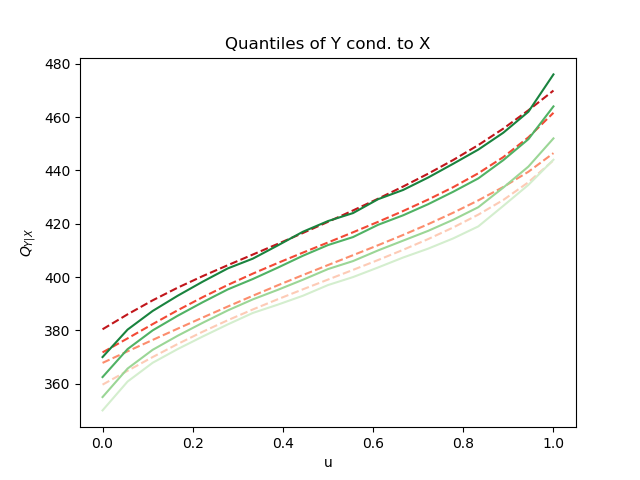}
  \caption*{Second dimension}
\end{minipage}
\caption{Comparison between one-dimensional VQR (regularized dual in dashed
red, with a ``soft" $\protect\phi$) and classical approach (green) with (i) $%
Y_1=$Weight (Left) or (ii) $Y_2=$Thigh circumference and $X=$(1, Height).
Quantiles are plotted for different height quantiles (10\%, 30\%, 60\%,
90\%). Regularization strengths are $\protect\varepsilon=0.1$. Chosen grid
size is $n=20$.}
\label{fig:weightthigh}
\end{figure}

\begin{figure}[]
\centering
\begin{minipage}{.25\textwidth}
  \centering
  \includegraphics[scale=0.2]{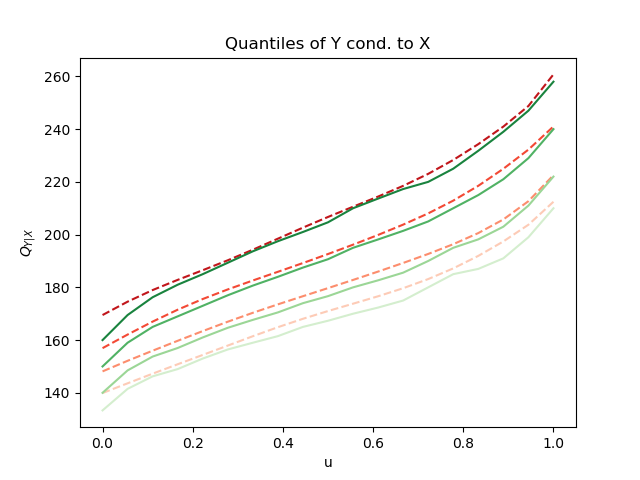}
  \\ $\varepsilon=0.05$
\end{minipage}\hspace*{0.01cm} 
\begin{minipage}{.25\textwidth}
  \centering
  \includegraphics[scale=0.2]{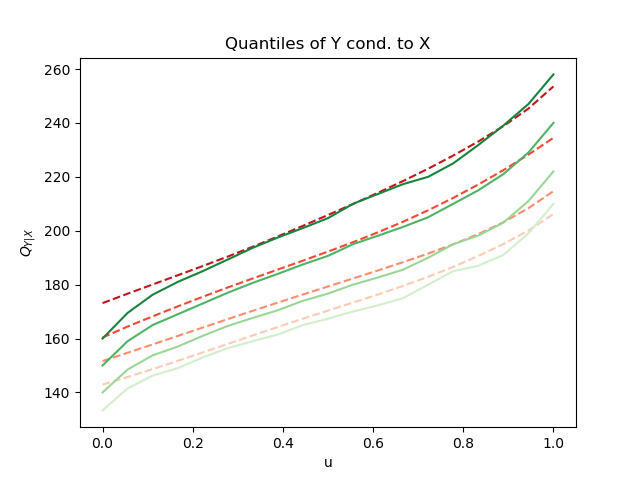}
  \\ $\varepsilon=0.1$
\end{minipage}\hspace*{0.01cm} 
\begin{minipage}{.25\textwidth}
  \centering
  \includegraphics[scale=0.2]{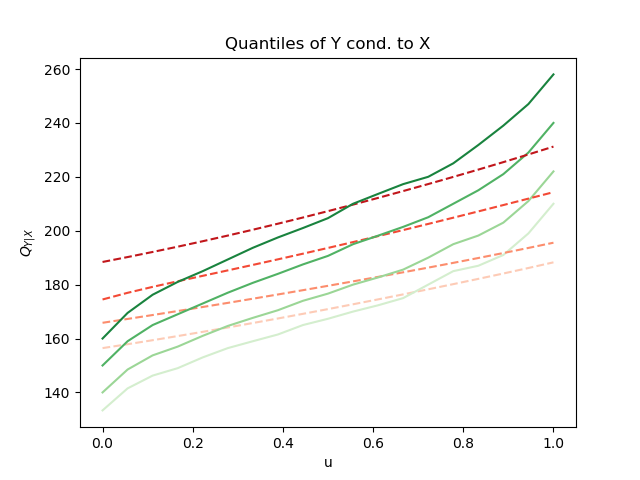}
  \\ $\varepsilon=0.5$
\end{minipage}\hspace*{0.01cm} 
\begin{minipage}{.25\textwidth}
  \centering
  \includegraphics[scale=0.2]{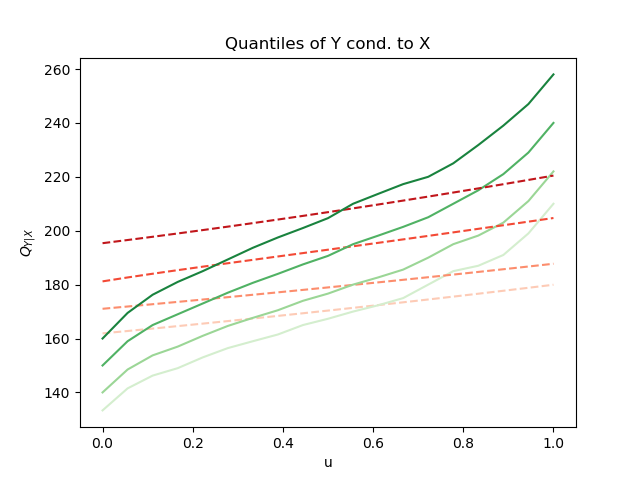}
  \\ $\varepsilon=1$
\end{minipage}
\caption{Regularized one-dimensional VQR, dual (dashed red) compared to
classical QR (green) with $Y_1=$Weight regressed on $X=$(1, Height), for
varying regularization strengths $\protect\varepsilon$. Quantiles are
plotted for different height quantiles (10\%, 30\%, 60\%, 90\%). Chosen grid
size is $n=20$.}
\label{fig:epsilon}
\end{figure}

{\textbf{One-dimensional VQR.}} We start by one-dimensional dependent
variables ($d=1$), namely \emph{Weight} ($Y_1$) and \emph{Thigh circumference%
} ($Y_2$), explained by $X=$(1, Height), to allow for comparison with
classical quantile regression of Koenker and Bassett (1978). Figure \ref%
{fig:weightthigh} displays results of our method compared to the classical
approach, for different height quantiles (10\%, 30\%, 60\%, 90\%). Figure %
\ref{fig:weightthigh} is computed with a ``soft" potential $\phi$ while
table \ref{tab:hardsoft} depicts the difference with its ``hard" counterpart
(see the beginning of section \ref{par:regul}). Figure \ref{fig:epsilon} and
Table \ref{tab:epsilon} detail the impact of regularization strength on
these quantiles.

\begin{table}[]
\vskip -0.15in \centering
\begin{tabular}{ccccc}
$\varepsilon$ & 0.05 & 0.1 & 0.5 & 1 \\ \hline
$\lvert\lvert Q_{soft}-Q_{hard} \rvert\rvert_2/\lvert\lvert Q_{soft}
\rvert\rvert_2$, $X=10\%$ & 3.8$\cdot 10^{-3}$ & 1.5$\cdot 10^{-2}$ & 6.7$%
\cdot 10^{-2}$ & 9.2$\cdot 10^{-2}$ \\ \hline
$\lvert\lvert Q_{soft}-Q_{hard} \rvert\rvert_2/\lvert\lvert Q_{soft}
\rvert\rvert_2$, $X=30\%$ & 6.8$\cdot 10^{-3}$ & 1.9$\cdot 10^{-2}$ & 7.0$%
\cdot 10^{-2}$ & 9.3$\cdot 10^{-2}$ \\ \hline
$\lvert\lvert Q_{soft}-Q_{hard} \rvert\rvert_2/\lvert\lvert Q_{soft}
\rvert\rvert_2$, $X=60\%$ & 1.2$\cdot 10^{-2}$ & 2.0$\cdot 10^{-2}$ & 6.9$%
\cdot 10^{-2}$ & 9.5$\cdot 10^{-2}$ \\ \hline
$\lvert\lvert Q_{soft}-Q_{hard} \rvert\rvert_2/\lvert\lvert Q_{soft}
\rvert\rvert_2$, $X=90\%$ & 1.6$\cdot 10^{-2}$ & 2.3$\cdot 10^{-2}$ & 6.8$%
\cdot 10^{-2}$ & 9.5$\cdot 10^{-2}$ \\ \hline
\end{tabular}
\caption{Relative error between one-dimensional VQR with a ``soft"
computation of $\protect\phi$ and its ``hard" counterpart, with $Y_1=$Weight
and $X=$(1, Height) for different height quantiles (10\%, 30\%, 60\%, 90\%),
depending on regularization strengths $\protect\varepsilon$. Chosen grid
size is $n=20$. }
\label{tab:hardsoft}
\end{table}

\begin{table}[]
\vskip -0.15in \centering
\begin{tabular}{ccccc}
$\varepsilon$ & 0.05 & 0.1 & 0.5 & 1 \\ \hline
$\lvert\lvert Q_{QR}-Q_{VQR} \rvert\rvert_2/\lvert\lvert Q_{QR}
\rvert\rvert_2$, $X=10\%$ & 9.8$\cdot 10^{-3}$ & 9.8$\cdot 10^{-3}$ & 2.8$%
\cdot 10^{-2}$ & 3.8$\cdot 10^{-2}$ \\ \hline
$\lvert\lvert Q_{QR}-Q_{VQR} \rvert\rvert_2/\lvert\lvert Q_{QR}
\rvert\rvert_2$, $X=30\%$ & 8.5$\cdot 10^{-3}$ & 1.1$\cdot 10^{-2}$ & 3.3$%
\cdot 10^{-2}$ & 4.3$\cdot 10^{-2}$ \\ \hline
$\lvert\lvert Q_{QR}-Q_{VQR} \rvert\rvert_2/\lvert\lvert Q_{QR}
\rvert\rvert_2$, $X=60\%$ & 7.7$\cdot 10^{-3}$ & 9.3$\cdot 10^{-3}$ & 3.1$%
\cdot 10^{-2}$ & 4.4$\cdot 10^{-2}$ \\ \hline
$\lvert\lvert Q_{QR}-Q_{VQR} \rvert\rvert_2/\lvert\lvert Q_{QR}
\rvert\rvert_2$, $X=90\%$ & 8.2$\cdot 10^{-3}$ & 1.0$\cdot 10^{-2}$ & 3.5$%
\cdot 10^{-2}$ & 4.9$\cdot 10^{-2}$ \\ \hline
\end{tabular}
\caption{Relative error between one-dimensional VQR and classical QR
approach with $Y_1=$Weight and $X=$(1, Height) for different height
quantiles (10\%, 30\%, 60\%, 90\%), depending on regularization strengths $%
\protect\varepsilon$. Chosen grid size is $n=20$. }
\label{tab:epsilon}
\end{table}

\begin{figure}[]
\centering
\includegraphics[scale=0.7]{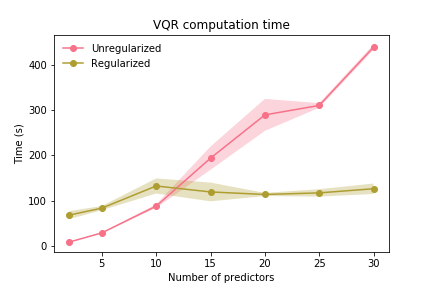}
\caption{Comparison of computational times between the unregularized case
(using Gurobi's barrier logging) and the regularized case, for a varying
number of predictors in 2D. In the latter, this time represents the time to
reach an error of $10^{-5}$ in $\left\Vert \cdot\right\Vert_2$ between two
iterates of the transport plan for $\protect\varepsilon=0.1$. Chosen grid
size is $n=10$ (per axis).}
\label{fig:times}
\end{figure}

\begin{figure}[]
\centering
\begin{tabular}{@{}c@{}c@{}c}
\begin{sideways}\parbox{.35\linewidth}{\centering  Small height
}\end{sideways} & \includegraphics[width=.5\linewidth]{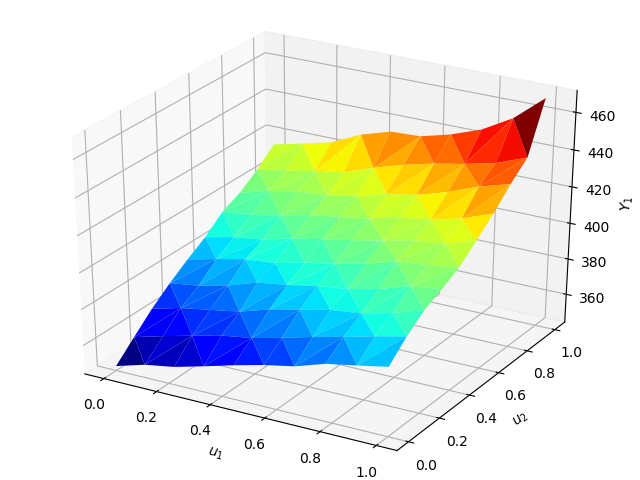} & %
\includegraphics[width=.5\linewidth]{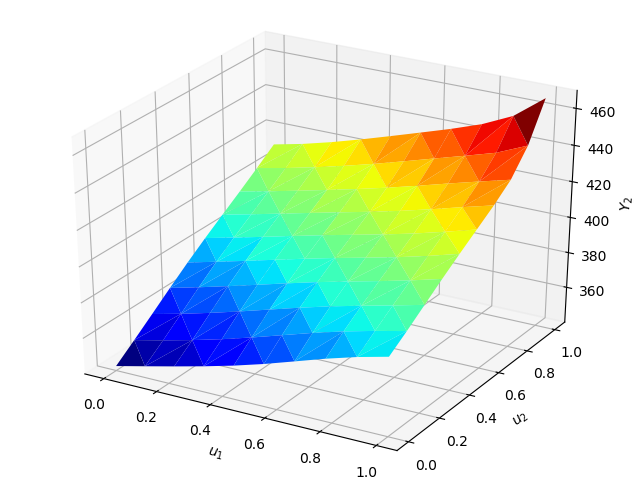} \\ 
\begin{sideways}\parbox{.35\linewidth}{\centering  Medium height
}\end{sideways} & \includegraphics[width=.5\linewidth]{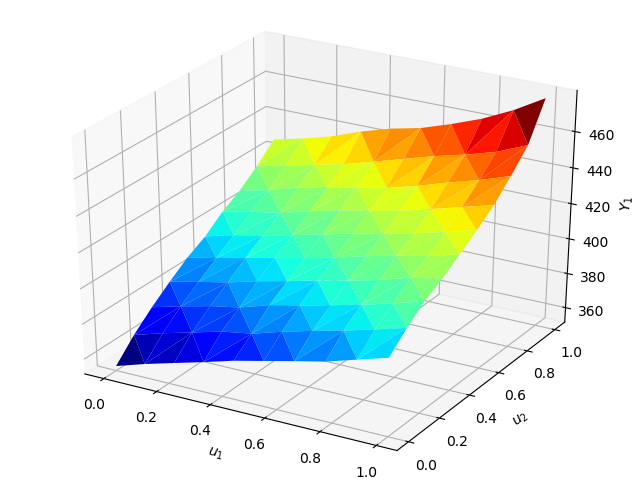} & %
\includegraphics[width=.5\linewidth]{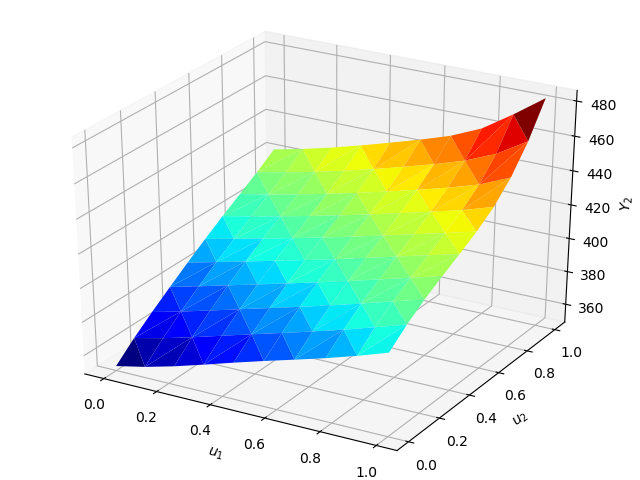} \\ 
\begin{sideways}\parbox{.35\linewidth}{\centering  Tall height
}\end{sideways} & \includegraphics[width=.5\linewidth]{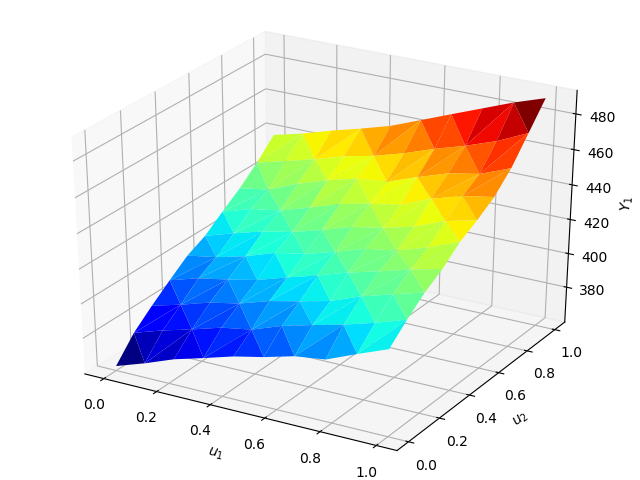} & %
\includegraphics[width=.5\linewidth]{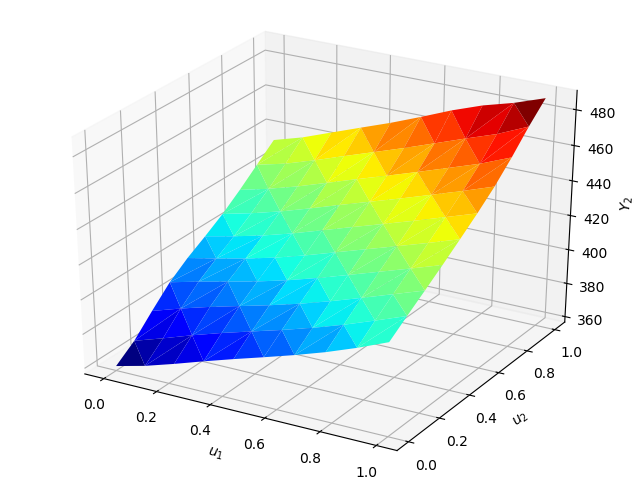} \\ 
& Unregularized & Regularized dual%
\end{tabular}
\caption{ Two-dimensional regularized quantile regression of $Y=$(Weight,
Thigh) explained by $X=$(1, Height). Quantiles of $Y_1$=Weight are plotted
for different height quantiles: 10\% (Bottom), 50\% (Middle) and 90\% (Top).
Chosen grid size is $n=10$ (per axis) and regularization strength $\protect%
\varepsilon=0.1$. }
\label{fig:multivariateWeight}
\end{figure}

\begin{figure}[]
\centering
\begin{tabular}{@{}c@{}c@{}c}
\begin{sideways}\parbox{.35\linewidth}{\centering  Small height
}\end{sideways} & \includegraphics[width=.5\linewidth]{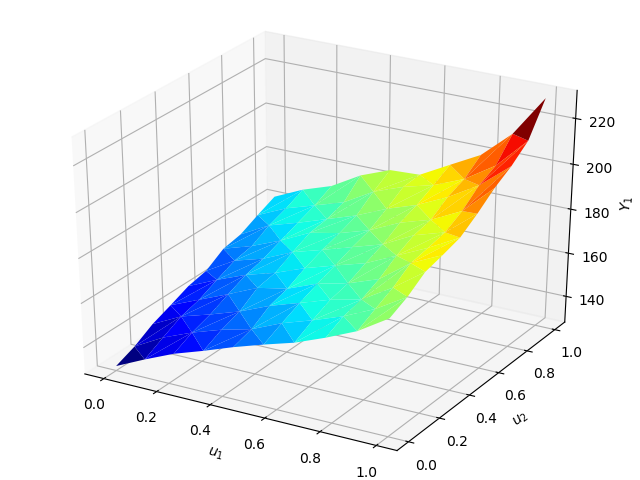} & %
\includegraphics[width=.5\linewidth]{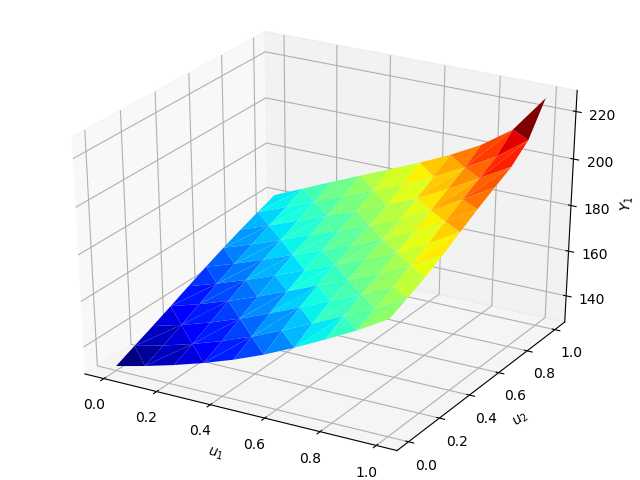} \\ 
\begin{sideways}\parbox{.35\linewidth}{\centering  Medium height
}\end{sideways} & \includegraphics[width=.5\linewidth]{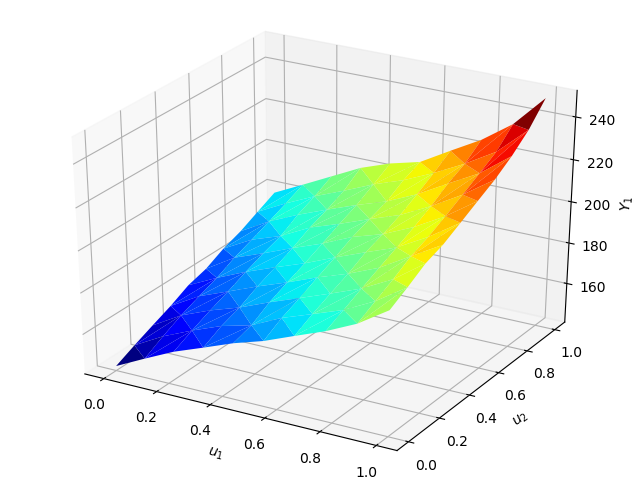} & %
\includegraphics[width=.5\linewidth]{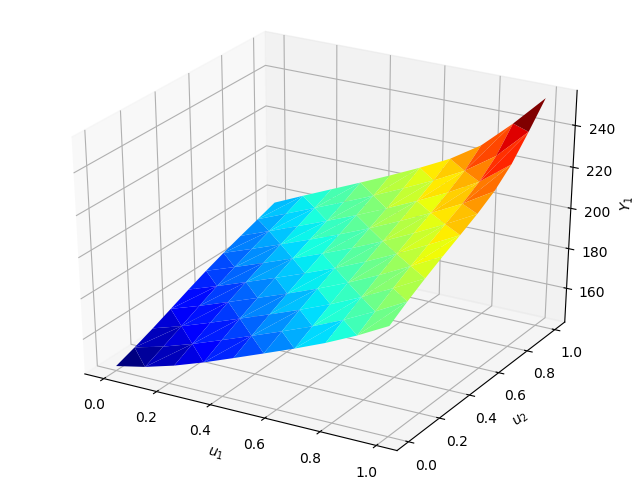} \\ 
\begin{sideways}\parbox{.35\linewidth}{\centering  Tall height
}\end{sideways} & \includegraphics[width=.5\linewidth]{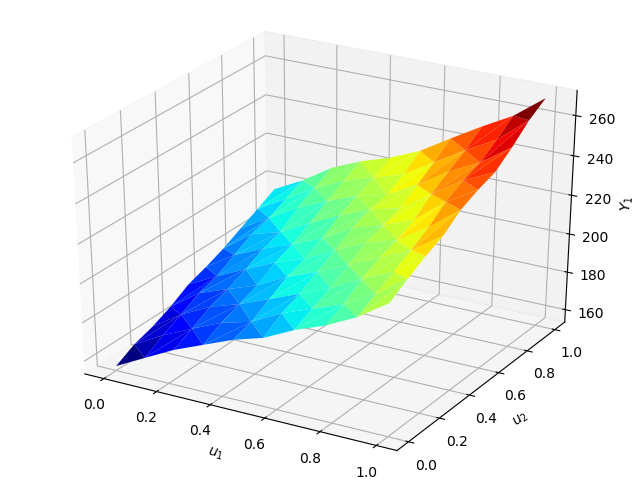} & %
\includegraphics[width=.5\linewidth]{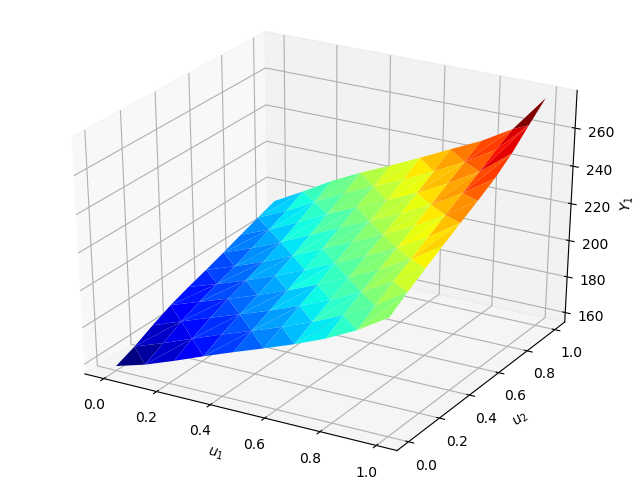} \\ 
& Unregularized & Regularized dual%
\end{tabular}
\caption{ Two-dimensional regularized quantile regression of $Y=$(Weight,
Thigh) explained by $X=$(1, Height). Quantiles of $Y_2$=Thigh are plotted
for different height quantiles: 10\% (Bottom), 50\% (Middle) and 90\% (Top).
Chosen grid size is $n=10$ (per axis) and regularization strength $\protect%
\varepsilon=0.1$. }
\label{fig:multivariateThigh}
\end{figure}

{\textbf{Multi-dimensional VQR.}} In contrast, multivariate quantile
regression explains the joint dependence $Y=(Y_1,Y_2)$ by $X=$(1,Height).
Figures \ref{fig:multivariateWeight} and \ref{fig:multivariateThigh} (each
corresponding to an explained component, either $Y_1$ or $Y_2$) depicts how
smoothing operates in higher dimension for different Height quantiles (10\%,
50\% and 90\%), compared to a previous unregularized approach Carlier et al.
(2016). Figure \ref{fig:times} details computational times in 2D using an
Intel(R) Core(TM) i7-7500U CPU 2.70GHz.

\clearpage

\section*{Appendix}

\subsection*{Proof of Lemma \protect\ref{suppC}}

Since $\mathbf{1}_{[0,t]}\in {\mathcal{C}}$, one obviously first has 
\begin{equation*}
\sup_{v\in {\mathcal{C}}}\int_{0}^{1}v(s)q(s)ds\geq \max_{t\in \lbrack
0,1]}\int_{0}^{t}q(s)ds=\max_{t\in \lbrack 0,1]}Q(t).
\end{equation*}%
Let us now prove the converse inequality, taking an arbitrary $v\in {%
\mathcal{C}}$. We first observe that $Q$ is absolutely continuous and that $%
v $ is of bounded variation (its derivative in the sense of distributions
being a bounded nonpositive measure which we denote by $\eta $), integrating
by parts and using the definition of ${\mathcal{C}}$ then give: 
\begin{equation*}
\begin{split}
\int_{0}^{1}v(s)q(s)ds& =-\int_{0}^{1}Q\eta +v(1^{-})Q(1) \\
& \leq (\max_{[0,1]}Q)\times (-\eta ([0,1])+v(1^{-})Q(1) \\
& =(\max_{[0,1]}Q)(v(0^{+})-v(1^{-}))+v(1^{-})Q(1) \\
& =(\max_{[0,1]}Q)v(0^{+})+(Q(1)-\max_{[0,1]}Q)v(1^{-}) \\
& \leq \max_{\lbrack 0,1]}Q.
\end{split}%
\end{equation*}

\end{document}